\documentclass[pra,preprint,showpacs]{revtex4} 
\usepackage{graphicx,bm,amsmath}
\begin{document}

\title{\bf Entanglement dynamics of two qubits under the influence of external kicks and Gaussian pulses} 
\author{Ferdi Altintas}
\author{Resul Eryigit}\email[email:]{resul@ibu.edu.tr}
\affiliation{Department of Physics, Abant Izzet Baysal University, Bolu, 14280-Turkey.}
\begin{abstract}
We have investigated the dynamics of entanglement between two spin-1/2 qubits that are subject to independent kick and Gaussian pulse type external magnetic fields analytically as well as numerically. Dyson time ordering effect on the dynamics is found to be important for the sequence of kicks. We show that "almost-steady" high entanglement can be created between two initially unentangled qubits by using carefully designed kick or pulse sequences.
\end{abstract}

\maketitle
\maketitle
\section{Introduction}
Control and manipulation of entanglement which is a quantifiable resource for quantum information tasks such as quantum computing~\cite{nc}, communication~\cite{bennett} and cryptography~\cite{ekert} have been studied along many directions in the last decade~\cite{hbbs,cmcfmgs,wl,levy,ms,zmf,wbsb,abliz,wang,sb}. Among these studies, systems that are modelled as 1-D Heisenberg chain with qubits of spin-1/2 particles as the main unit are one of the prototypical examples~\cite{iabdlss}. For such systems the control of entanglement between the qubits can be manipulated with the help of various type external magnetic fields~\cite{hbbs,cmcfmgs,wl,levy,ms,zmf,wbsb,abliz,wang,sb}. In particular, Heule {\it et al.} investigated the feasibility of local operator control in arrays of interacting qubits modelled as isotropic Heisenberg spin chains~\cite{hbbs}. Along similar lines, Caneva {\it et al.} explored optimal quantum control by  appropriate pulses to affect the required transformations by numerical Krotov algorithm~\cite{cmcfmgs}. Wu {\it et al.} showed that one qubit gates can be constructed with global magnetic fields and controllable Heisenberg exchange interactions~\cite{wl}. Levy demonstrated a scheme that uses pairs of spin-1/2 particles to form logic qubits and Heisenberg exchange only to produce all gate operations~\cite{levy}. Malinovsky and Sola studied phase control of entanglement in two qubit systems and showed that by changing the relative phase of control pulses, one can control entanglement at will~\cite{ms}. Sadiek {\it et al.} studied the control and manipulation of entanglement evolution for a two qubit system coupled through XYZ Heisenberg interaction influenced by a time-varying external field~\cite{zmf}. Wang {\it et al.} demonstrated that near perfect entanglement can be obtained by applying a magnetic field on a single spin of an isotropic Heisenberg chain of length $N$~\cite{wbsb}.  Abliz {\it et al.} studied the entanglement dynamics for a two-qubit Heisenberg XXZ model effected by population relaxation in the presence of various types of magnetic fields and showed that it is possible to produce, control and modulate high entanglement with the help of time-dependent external fields despite the existence of dissipation~\cite{abliz}.

For a general time-dependent external field, time-ordering effects might be important and the dynamics cannot be found analytically. Most of the aforementioned studies employ numerical methods to investigate the entanglement control~\cite{hbbs,cmcfmgs,ms,zmf,wbsb}. Although the numerical methods are fast and reliable, analytic solutions provide a more clear picture of the physics behind the dynamics. For a single qubit, the time evolution of populations and coherence under the influence of external field in the form of Gaussian pulse or a delta function kick was investigated by Kaplan {\it et al.} and Shakov {\it et al.}~\cite{kaplan,shakov}. The fast pulse or kick is defined based on the relation between the energy splitting of the qubit $\Delta E$  and the duration of the pulse $\tau$; if $\Delta E\tau<<1$ then the pulse is called a kick~\cite{kaplan,shakov}. Many experimental and theoretical studies have been carried out to investigate the dynamics of two level quantum systems under the influence of a single or a series of kicks for quantum gates~\cite{jhm,vsbysc}, NMR~\cite{slichter}, excitation of electronic states in molecules~\cite{krgt}, chemical reactions~\cite{swr} and quantum computing~\cite{pk}. However, there is no study of control of entanglement between two qubits by using fast pulses, to the best of our knowledge.

Fast pulses provide an efficient way of full population transfer in a qubit~\cite{kaplan,shakov} and because of that are expected to be an important way of controlling the entanglement. From this point, in the present study, we consider two qubits with Heisenberg XXX-type interaction. Each qubit is under the influence of a local time-dependent magnetic field that acts in the $z$-direction. We consider one, two, three and four kicks as well as Gaussian pulse sequences and their effect on the dynamics of entanglement between the qubits. We show that entanglement can be controlled by a careful design of the sequence of kicks.

The organization of this paper is as follows: In Sec.~\ref{secbasic}, we introduce the model and basic formalism necessary to solve time evolution exactly. In Sec.~\ref{nto}, we discuss time ordering effect on the dynamics. In Sec.~\ref{concurrence}, Wootters concurrence as an entanglement measure is briefly introduced. In Sec.~\ref{seckicked}, the analytic entanglement dynamics of kicked qubits in the presence of time ordering is discussed by choosing  single and multiple~(up to four) kicks. In Sec.~\ref{gaussian}, the effect of finite pulse width on the entanglement dynamics is studied numerically by choosing a Gaussian pulse or pulse sequence as an external field. We conclude as a summary of the important results in Sec.~\ref{conc}. 

\section{The model and basic formulation}
\label{secbasic}
In this paper, we consider two Heisenberg XXX coupled qubits in a time-dependent external magnetic field acting in the $z$-direction. The typical time-dependent Hamiltonian for this system may be expressed as~\cite{zmf}~(we set $\hbar=1$):
\begin{eqnarray}
\label{totalH}
\hat{H}(t) &=& \hat{H}_0+\hat{H}_{int}(t),
\end{eqnarray}
where
\begin{eqnarray}
\label{individualH}
\hat{H}_0 &=& J \displaystyle\sum_{i=x,y,z}\hat{\sigma}_{i}^1 \hat{\sigma}_{i}^2,\nonumber\\
\hat{H}_{int}(t)&=&-\displaystyle\sum_{i=1}^2 B_{z}^i(t)\hat{\sigma}_{z}^i,
\end{eqnarray}
where $\hat{\sigma}_i^{1,2}~(i=x,y,z)$ are the usual Pauli spin matrices, $J$ is the qubit-qubit interaction strength and $B_z^{1}(t)$ and $B_z^{2}(t)$ are the time-dependent magnetic fields acting on  qubit $1$  and $2$, respectively. It should be noted that the qubit-qubit interaction term in Eq.~(\ref{totalH}) is given by $\hat{H}_0$ which is constant in time and the time-dependent part of the total Hamiltonian is called $\hat{H}_{int}(t)$ which describes the qubit-magnetic field interaction and assumed to be a single real function of $t$.

The most general form of an initial pure state of the two-qubit system is $\left|\Psi(0)\right\rangle=a_1(0)\left|11 \right\rangle +a_2(0)\left|10\right\rangle+a_3(0)\left|01\right\rangle +a_4(0)\left|00\right\rangle$, where $a_i(0)~(i=1,2,3,4)$ are complex numbers with $\displaystyle\sum_{i=1}^4|a_i(0)|^2=1$, then the probability amplitudes evolve in time under Hamiltonian~(\ref{totalH}) according to Schr\"{o}dinger equation as:
\begin{eqnarray}
\label{hamiltonian}
i{d \over dt}
\left [\begin{array}{c} a_1(t) \\ a_2(t) \\ a_3(t) \\ a_4(t) \end{array} \right]
 = \left[
\begin{array}{cccc} J-B_T(t) & 0 & 0 & 0 \\ 0 & -J+\Delta B(t) & 2 J & 0 \\ 0 & 2 J & -J-\Delta B(t) & 0 \\ 0 & 0 & 0& J+B_T(t)\end{array}\right]
\left [ \begin{array}{c} a_1(t) \\ a_2(t) \\ a_3(t) \\ a_4(t)\end{array}\right],\nonumber\\
\end{eqnarray}
where $\Delta B(t)=B_z^2(t)-B_z^1(t)$ and $B_T(t)=B_z^1(t)+B_z^2(t)$. The formal solution of Eq.~(\ref{hamiltonian}) may be written in terms of the time evolution matrix $\hat{U}(t)$ as
\begin{eqnarray}
\label{amps}
\left [ \begin{array}{c} a_1(t) \\ a_2(t) \\ a_3(t) \\ a_4(t) \end{array} \right] 
 = \hat U(t)
\left [ \begin{array}{c} a_1(0) \\ a_2(0) \\ a_3(0) \\ a_4(0) \end{array} \right].
\end{eqnarray}
The evolution operator for the general time-dependent Hamiltonian of two qubits is not easy to obtain analytically; a number of systematic procedures are obtained in Refs~\cite{rau} and~\cite{rsu} based on dynamical groups of the system when time-ordering is not important. Here the time evolution operator $\hat{U}(t)$ may be expressed as:
\begin{eqnarray}
\label{U}
  \hat U(t)&=&\hat{T}e^{-i\int^t_0\hat{H}(t')dt'}=\hat{T}e^{-i\int^t_0\left(\hat{H}_0+\hat{H}_{int}(t')\right)dt'}\nonumber\\
	&=&\hat{T}\sum_{n=0}^\infty \frac{(-i)^n}{n!}\int_0^t\hat{H}(t_n)dt_n ...\int_0^t\hat{H}(t_2)dt_2\int_0^t\hat{H}(t_1) dt_1.
\end{eqnarray}
The only non-trivial time dependence in $\hat{U}(t)$ arises from time-dependent $\hat H(t)$ and time ordering $\hat{T}$. The Dyson time ordering operator $\hat{T}$ specifies that $\hat{H}(t_i) \hat{H}(t_j)$ is properly ordered~\cite{kaplan,shakov,dysont}: 
\begin{eqnarray}
\label{Tpro}
\hat{T}\hat{H}(t_i)\hat{H}(t_j)=\hat{H}(t_i)\hat{H}(t_j)+\theta(t_j-t_i)\left[\hat{H}(t_j),\hat{H}(t_i)\right],
\end{eqnarray}
where $\theta(t_j-t_i)$ is the Heaviside step function whose value is zero if $(t_j-t_i)$ is negative and one if $(t_j-t_i)$ is positive. It should be noted that time ordering imposes a connection between the effects of $\hat{H}(t_i)$ and $\hat{H}(t_j)$ and gives rise to observable, non-local, time ordering effects when $\left [\hat{H}(t_j),\hat{H}(t_i)\right]\ne0$~\cite{zlt,mbhgm}.

\section{Time ordering}\label{nto}
\label{sectimeorder}
If one takes $\hat{T}=1$ in Eq.~(\ref{U}) to obtain $\hat{U}(t)$, the time evolution is said to contain no time ordering. So the difference between the result obtained by an exact treatment of $\hat{T}$ in Eq.~(\ref{U}) and $\hat{T}\rightarrow1$ is called the effect of time ordering on the dynamics~\cite{kaplan,shakov}. One should note that removing time ordering as $\hat{T}\rightarrow1$ corresponds to the zeroth order term in an eikonal-like, Magnus expansion in the commutator terms~\cite{magnus}.

\subsection{Limit of no time ordering}
Replacing $\hat{T}$ with $1$ in Eq.~(\ref{U}), in the Schr\"{o}dinger picture we
have,
\begin{eqnarray}
\label{U0}
\hat{U}(t)&=&\hat{T}e^{-i\int^t_0\hat{H}(t')dt'}\to\sum_{n=0}^\infty\frac{(-i)^n}{n!} \left[\int_0^t\hat{H}(t')dt'\right]^n\nonumber\\
&=&\sum_{n=0}^\infty\frac{(-i)^n}{n!}\left[\hat{H}_0t+\int_0^t\hat{H}_{int}(t') dt'\right]^n= \sum_{n=0}^\infty\frac{(-i)^n}{n!}\left[\left(\hat{H}_0+\hat{\bar{H}}_{int}\right)t\right]^n\nonumber\\
&=&e^{-i\hat{\bar{H}}t}=\hat{U}^0(t),
\end{eqnarray}
where
$\hat{\bar H}_{int}t=\int_0^t\hat H_{int}(t')dt'$ is the averaged interaction field, and $\hat{\bar H}=\hat H_0+\hat{\bar H}_{int}$ and $[\hat H_0,\hat{\bar H}_{int}]$ terms are non-zero. By expanding in powers of $[\hat{H}(t''), \hat{H}(t')]$, it is
straightforward to show that to leading order in $\hat{H}_{int}(t)$ and $\hat{H}_0$, the
time ordering effect is given by
\begin{eqnarray}
\label{expanding}
\hat{U} - \hat{U}^0\simeq -\frac{1}{2}\int_0^t dt''\int_0^{t''}dt'\left[\hat{H}(t''),\hat{H}(t')\right]=-\frac{1}{2}\left[\hat{H}_0,
\hat{H}_{int}^0\right]\int_0^tdt'(t - 2t')f(t'),\nonumber\\
\end{eqnarray}
where $\hat{H}_{int}(t')=\hat{H}_{int}^0f(t')$.  This leading term disappears if the pulse centroid $T_k=t/2$ and $f(t')$ is symmetric about $T_k$.  Furthermore, $\hat{U}-\hat{U}^0$ vanishes identically in the special cases of $H_{int}(t')=0, H_{int}(t')=\bar{H}_{int}$~\cite{kaplan,shakov}. Also the commutator $\left[\hat{H}(t''),\hat{H}(t')\right]$~(i.e., the time ordering effect) vanishes for $B_{z}^1(t)=B_{z}^2(t)$ or $J=0$ because by using the total Hamiltonian~(\ref{totalH}) we have
\begin{eqnarray}
\label{com}
[\hat{H}(t''),\hat{H}(t')]=2iJ\left(\left(B_{z}^1(t')-B_{z}^2(t')\right)-\left(B_{z}^1(t'')-B_{z}^2(t'')\right)\right)(\hat{\sigma}_y^1\hat{\sigma}_x^2-\hat{\sigma}_x^1\hat{\sigma}_y^2),\nonumber\\
\end{eqnarray}
which vanishes when either the time-dependent magnetic fields on qubits 1 and 2 are equal or the qubit-qubit interaction is neglected. 

In general there is no simple analytic form for the exact result $\hat{U}(t)$, except for special cases~\cite{zmf,kaplan,shakov}. For the result without time ordering with averaged magnetic fields $\bar{B}_{z}^1t=\int_0^t B_{z}^1(t')dt'=\alpha$ and $\bar{B}_{z}^2t=\int_0^tB_{z}^2(t')dt'=\beta$, the time evolution matrix $\hat{U}^0(t)$ in Eq.~(\ref{U0}) can be easily calculated as: 
\begin{eqnarray}
\label{u0pulse}
\hat{U}^0(t)&=&e^{-i(\hat H_0t+\hat{\bar{H}}_{int}t)}\nonumber\\
&=&\left[\begin{array}{cccc} y_{1} y^{*}  & 0 & 0  & 0 \\ 0  & y (u+iv) & y (-w+i z)  & 0 \\ 0  & y (w+i z) & y (u-iv)  & 0 \\ 0  & 0 & 0  & y_{1}^{*} y^{*}
\end{array} \right],
\end{eqnarray}
where
\begin{eqnarray}
\label{u0parameters}
y&=&e^{i J t},\nonumber\\
y_1&=&e^{i(\alpha+\beta)},\nonumber\\
u&=&\cos \left(\Gamma\right),\nonumber\\
v&=&\frac{(\alpha-\beta)}{\Gamma}\sin\left(\Gamma\right),\nonumber\\
w&=&0,\nonumber\\
z&=&-\frac{2 J t}{\Gamma}\sin\left(\Gamma\right),
\end{eqnarray}
where $\Gamma =\sqrt{4 J^2 t^2+(\alpha-\beta)^2}$ and $\alpha$ and $\beta$ are called the integrated magnetic strengths associated with the magnetic fields acting on qubit $1$ and $2$, respectively.

Similarly, the time evolution matrix without time ordering in interaction picture can be studied~\cite{kaplan,shakov}. However in Refs.~\cite{kaplan} and~\cite{shakov}, it was shown that the  occupation probabilities for a kicked qubit for the dynamics in the limit $\hat{T}\rightarrow 1$ depend on the chosen picture. On the other hand, the exact results~(the results including time ordering) are independent of the chosen picture, as expected.  Thus in the next sections, we will discuss the entanglement dynamics of kicked qubits by using exact time ordered results, and we will work in Schr\"{o}dinger representation. The matrix in Eq.~(\ref{u0pulse}) will be used to check the correctness of our results, because as mentioned before, when $B_z^1(t)=B_z^2(t)$ or $J=0$, the time ordering effect defined as $\hat{U}^K(t)-\hat{U}^0(t)$ vanishes.

\section{Measure of entanglement}
\label{concurrence}
For a pair of qubits, Wootters concurrence can be used as a measure of entanglement~\cite{wcon}. The concurrence function varies from $C=0$ for a separable state to $C=1$ for a maximally entangled state. To calculate the concurrence function one needs to evaluate the matrix 
\begin{eqnarray}
\label{rhotrans}
\hat{R}=\hat{\rho}(t) (\hat{\sigma}_y\otimes\hat{\sigma}_y)\hat{\rho}^*(t)(\hat{\sigma}_y\otimes\hat{\sigma}_y),
\end{eqnarray}
where $\hat{\rho}(t)$ is the density matrix of the system and $\hat{\rho}^*(t)$ is its complex conjugate. Then the concurrence is defined as 
\begin{eqnarray}
\label{con}
C(\hat{\rho})=\max\{0,\lambda_1-\lambda_2-\lambda_3-\lambda_4\},
\end{eqnarray}
where $\lambda_1,\lambda_2,\lambda_3$ and $\lambda_4$ are the positive square roots of the eigenvalues of $\hat{R}$ in descending order.

It should be noted that due to the discrete symmetry (conservation of parity under flipping of the $\hat{\sigma}_i^j$, $i=x,y,z$ and $j=1,2$, i.e. when $\hat{\sigma}_i^j\rightarrow-\hat{\sigma}_i^j$)  of the total Hamiltonian~(\ref{totalH}), the states $\left|\Phi\right\rangle=a_2\left|10 \right\rangle+a_3\left|01\right\rangle$ and $\left|\Psi\right\rangle=a_1\left|11\right\rangle+a_4\left|00\right\rangle$ can never get mixed in time due to that symmetry~\cite{zmf}, as can be seen from Eq.~(\ref{hamiltonian}). Thus we consider the time evolution of the concurrence of these states individually. The concurrence function for a pure state $\left|\Phi(t) \right\rangle=a_2(t) \left|10 \right\rangle+a_3(t) \left|01 \right\rangle$ with density matrix $\hat{\rho}(t)=\left|\Phi(t)\right\rangle\left\langle\Phi(t)\right|$ is given by
\begin{eqnarray}
\label{con1}
C(\hat{\rho})=\max\{0,2 \left|a_2(t) a_3(t)\right|\}.
\end{eqnarray}
Similarly, for the pure state $\left|\Psi(t) \right\rangle=a_1(t) \left|11 \right\rangle+a_4(t)\left|00 \right\rangle$ with density matrix $\hat{\rho}(t)=\left|\Psi(t)\right\rangle\left\langle\Psi(t)\right|$, the concurrence function reads as
\begin{eqnarray}
\label{con2}
C(\hat{\rho})=\max\{0,2 \left|a_1(t) a_4(t)\right|\},
\end{eqnarray}
where according to Eq.~(\ref{amps}), the time-dependent coefficients read
\begin{eqnarray}
\label{a1234}
a_i(t)&=&\displaystyle\sum_{j=1}^{4}U_{ij}(t)a_j(0),
\end{eqnarray}
where $U_{ij}(t)~(i,j=1,2,3,4)$ are the matrix elements of $\hat{U}(t)$.
\section{Entanglement dynamics of kicked qubits}
\label{seckicked}
In this part, we will examine the entanglement dynamics of kicked qubits by taking into account the time ordering effects for the initially pure separable $\left|\Psi(0)\right\rangle=\left|01\right\rangle$ and maximally entangled $\left|\Phi(0)\right\rangle=\frac{1}{\sqrt{2}}(\left|10 \right\rangle +\left|01\right\rangle)$ Bell states. We will work in the Schr\"{o}dinger picture and present analytic expressions for the time evolution operator of the two-qubit system for a single kick as well as a positive followed by a negative kick and a sequence of two, three and four equally distanced kicks.  For all kick sequences we consider two integrated magnetic strength regimes: $\alpha=2\beta$ and $\alpha=3\beta$ and for convenience we shall set $J=1$ and $\beta=1$. From those results we will use Eqs.~(\ref{con1}) and~(\ref{a1234}) to analyze and discuss the time evolution of the entanglement between the qubits.

We have also considered the other Bell states $\frac{1}{\sqrt{2}}(\left|11\right\rangle\pm\left|00\right\rangle)$ and $\frac{1}{\sqrt{2}}(\left|10\right\rangle-\left|01\right\rangle)$ as well as separable states of two qubits as $\left|11\right\rangle$, $\left|00\right\rangle$, $\left|10\right\rangle$. Under the influence of $\hat{H}(t)$ of Eq.~(\ref{totalH}), the states of the type $\left|00\right\rangle$ and $\left|11\right\rangle$ remain separable, while the entanglement of $\frac{1}{\sqrt{2}}(\left|11\right\rangle\pm\left|00\right\rangle)$ does not change with time. These may be checked using Eq.~(\ref{con2}) and the solution of the expansion coefficients $a_1(t)$ and $a_4(t)$ in Eq.~(\ref{hamiltonian}) for the considered initial states.  The dynamics of $\frac{1}{\sqrt{2}}(\left|10\right\rangle-\left|01\right\rangle)$ is same as that of $\frac{1}{\sqrt{2}}(\left|10\right\rangle+\left|01\right\rangle)$ and that of  $\left|01\right\rangle$ is same as $\left|10\right\rangle$ that are noted after specifying the propagators for kicked qubits and by using Eq.~(\ref{con1}) and Eq.~(\ref{a1234}). So, we consider only $\left|\Psi(0)\right\rangle=\left|01\right\rangle$ and $\left|\Phi(0)\right\rangle=\frac{1}{\sqrt{2}}(\left|10 \right\rangle +\left|01\right\rangle)$ as initial states.

One point we want to emphasize that before the field is active, the propagator is equal to $e^{-i\hat{H}_0t}$ and given by Eq.~(\ref{freeevolution}). Based on Eqs.~(\ref{con1}),~(\ref{a1234}) and~(\ref{freeevolution}), the concurrence of the initial state $\left|\Phi(0)\right\rangle=\frac{1}{\sqrt{2}}(\left|10\right\rangle+\left|01\right\rangle)$ is equal to $1$, while the concurrence of $\left|\Psi(0)\right\rangle=\left|01 \right\rangle$ is $|\sin(4Jt)|$ at time $t$ before the kick. Note that the entanglement dynamics  for the initial state $\left|\Phi(0)\right\rangle=\frac{1}{\sqrt{2}}(\left|10 \right\rangle +\left|01\right\rangle)$ is unperturbed by the qubit-qubit interaction in the absence of external field, while qubit-qubit interaction creates a high degree of entanglement that oscillates between $0$ and $1$ if the qubits are initially prepared in $\left|\Psi(0)\right\rangle=\left|01\right\rangle$ state.
\subsection{Single kick}
\begin{figure}[!hbt]\centering
\label{fig_singlekick}
{\scalebox{0.3}{\includegraphics{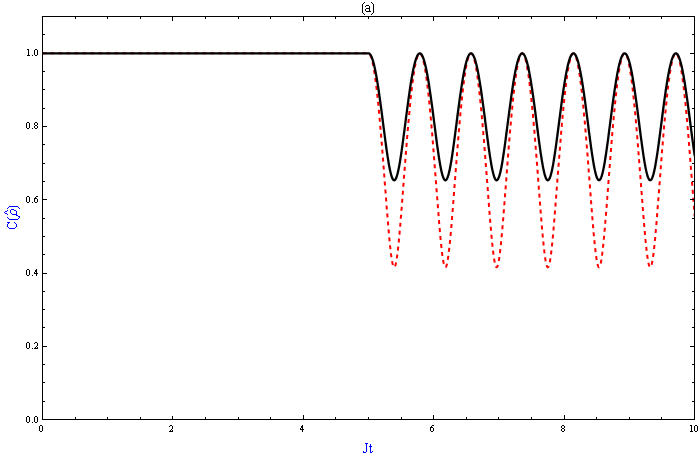}}}
{\scalebox{0.3}{\includegraphics{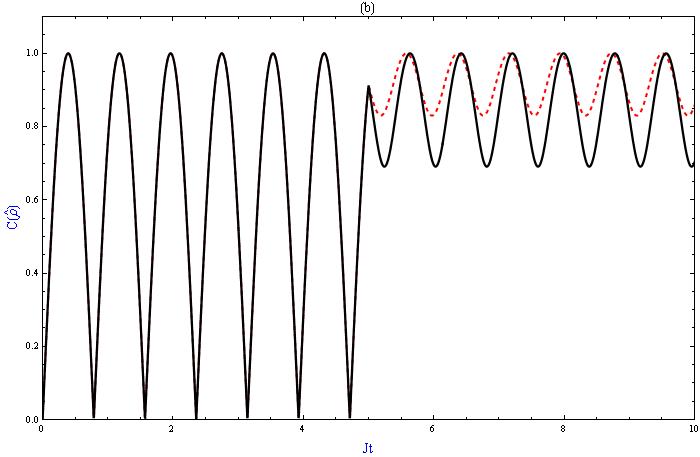}}}
\caption{Concurrence as a function of dimensionless time, $Jt$, for an ideal positive kick applied at $T_1=5$ for the initial pure states $\left|\Phi(0)\right\rangle=\frac{1}{\sqrt{2}}(\left|10\right\rangle+\left|01\right\rangle)$~(a) and $\left|\Psi(0)\right\rangle=\left|01\right\rangle$~(b). The dashed lines correspond to $\alpha=2 \beta$ and the solid lines to $\alpha=3 \beta$.}
\end{figure}
Here we consider two qubits whose states coupled by an interaction field which can be
expressed as a sudden "kick" at $t = T_1$, namely $B_z^1(t)=\alpha \delta(t-T_1),B_z^2(t)=\beta \delta(t-T_1)$.  
For such a kick the integration over time is trivial and the
time evolution matrix in Eq.~(\ref{U}) becomes~\cite{kaplan}
\begin{eqnarray}
\label{U^K}
\hat U^K(t) = e^{-i\hat{H}_0(t-T_1)}e^{-i\int_{T_1-\epsilon}^{T_1+\epsilon}\hat{H}_{int}(t') dt' }
e^{-i\hat{H}_0T_1},
\end{eqnarray}
in the same form as Eq.~(\ref{u0pulse}) with elements
\begin{eqnarray}
\label{U^Kpara}
y&=&e^{iJt},\nonumber\\
y_{1}&=&e^{i(\alpha+\beta)},\nonumber\\
u&=&\cos \left(2 J t\right) \cos(\alpha-\beta),\nonumber\\
v&=&\cos\left(2J(t-2T_1)\right)\sin(\alpha-\beta),\nonumber\\
w&=&\sin \left(2J(t-2T_1)\right)\sin(\alpha-\beta ),\nonumber\\
z&=&-\sin\left(2Jt\right)\cos(\alpha-\beta),
\end{eqnarray}
for $t > T_1$. The propagator without time ordering is given in Eq.~(\ref{u0pulse}) and as explained before when $\alpha=\beta$ or $J=0$, the time ordering effect, $\hat{U}^K(t)-\hat{U}^0(t)$, vanishes after the field is active.

By inserting Eq.~(\ref{U^Kpara}) into Eqs.~(\ref{con1}) and~(\ref{a1234}) for the initial states $\left|\Phi(0)\right\rangle=\frac{1}{\sqrt{2}}(\left|10\right\rangle+\left|01\right\rangle)$ and $\left|\Psi(0)\right\rangle=\left|01\right\rangle$, one can obtain the analytic expressions of the concurrence functions after the kick~($t>T_1$). For the maximally entangled state, the concurrence is given as:
\begin{eqnarray}\label{skcb}
C(\hat{\rho})=\max\{0,\left|\cos^2(\Delta)+e^{8iJ(t-T_1)}\sin^2(\Delta)\right|\},
\end{eqnarray}
while for $\left|\Psi(0)\right\rangle=\left|01\right\rangle$ the concurrence after $t=T_1$ can be obtained as
\begin{eqnarray}\label{skcs}
C(\hat{\rho})=2\max\{0,|\Lambda|\},
\end{eqnarray}
where $\Lambda=\left(\cos(2Jt)\cos(\Delta)-i\cos(\zeta)\sin(\Delta)\right)\left(i\cos(\Delta)\sin(2Jt)+\sin(\zeta)\sin(\Delta)\right)$, $\Delta=\alpha-\beta$ and $\zeta=2J(t-2T_1)$.

The dynamics of concurrence for the initial Bell state and the separable state for the single kick which are given by Eqs.~(\ref{skcb}) and~(\ref{skcs}) for $t>T_1$ are displayed in Fig.~1(a) and~(b), respectively. The effect of the kick on the entanglement is pronounced for both initial states; the concurrence of the Bell state starts oscillating with an amplitude that depends on the ratio of the integrated magnetic strength of the external fields on qubit 1 and 2~(i.e., $\alpha$ and $\beta$, respectively). The effect of the kick on the system initially in separable state, $\left|01\right\rangle$, is similar with the exception that the concurrence variation amplitudes get lower after the kick. One should also note that $C(\hat{\rho})$ of the initial Bell state is independent of $J$ before the kick, while the frequency of its time-dependence after the kick is proportional to the qubit-qubit interaction strength $J$, as can be seen from Fig.~1(a) and the analytic expression Eq.~(\ref{U^Kpara}).

\subsection{A positive followed by a negative kick}
\label{secmultkick}
The propagator for a sequence of  either identical or non-identical kicks can be easily obtained by
multiplication of several matrices of the form of Eq.~(\ref{U^K})~\cite{kaplan,shakov}. For example, one may consider a sequence of two kicks of opposite sign at times $t=T_1$ and $t=T_2$, namely, $B_z^1(t)=\alpha\left(\delta(t-T_1)-\delta(t-T_2)\right)$, $B_z^2(t)=\beta\left(\delta(t-T_1)-\delta(t-T_2)\right)$.  Following the procedure given in Eq.~(\ref{U^K}), one obtains the time evolution matrix for $t > T_2$ as~\cite{kaplan}:
\begin{eqnarray}
\label{kickakick}
\hat{U}^{K}(t)=e^{-i\hat{H}_0(t-T_2)}e^{-i\int_{T_2-\epsilon}^{T_2 +\epsilon}\hat{H}_{int}(t') dt'}e^{-i\hat{H}_0(T_2-T_1)}e^{-i\int_{T_1-\epsilon}^{T_1+\epsilon}\hat{H}_{int}(t')dt'} e^{-i\hat{H}_0T_1},\nonumber\\
\end{eqnarray}
where the elements of Eq.~(\ref{kickakick}) is the same form as Eq.~(\ref{u0pulse}) with parameters 
\begin{eqnarray}
\label{kickakickpara}
y&=&e^{iJt},\nonumber\\
y_{1}&=&1,\nonumber\\
u&=&\cos\left(2Jt\right)\cos(\Delta)^2+\cos\left(2J(t-2 T_s)\right)\sin(\Delta)^2,\nonumber\\
v&=&\left(\cos(\zeta_1)-\cos(\zeta_2)\right)\sin(\Delta)\cos(\Delta),\nonumber\\
w&=&\left(\sin(\zeta_1)-\sin(\zeta_2)\right)\sin (\Delta)\cos(\Delta),\nonumber\\
z&=&-\sin\left(2Jt\right)\cos(\Delta)^2-\sin\left(2J(t-2 T_s)\right)\sin(\Delta)^2,
\end{eqnarray}
where $\zeta_i=2J(t-2 T_i),\Delta=(\alpha-\beta)$ and $T_s = T_2 - T_1$. For this case, the time evolution matrix without time ordering is given by
\begin{eqnarray}
\label{freeevolution}
\hat{U}^{0}(t)&=&e^{-i\hat{H}_0 t}\nonumber\\
&=& \left [\begin{array}{cccc}
  e^{-iJt}  & 0 & 0  & 0 \\
  0  & e^{iJt}\cos(2Jt) &-i e^{iJt} \sin(2Jt)  & 0 \\ 
  0  & -i e^{iJt}\sin(2Jt)  & e^{iJt}\cos(2Jt) & 0  \\
  0  & 0 & 0  & e^{-iJt}\end{array} \right],
\end{eqnarray}
because for a positive kick followed by a negative kick the averaged interaction Hamiltonian, $\hat{\bar{H}}_{int}t=0$. For the cases $J=0$, or $T_s=0$, or $\alpha=\beta$, the time ordering effect defined as $\hat{U}^K(t)-\hat{U}^0(t)$ goes to zero, as expected.
\begin{figure}[!hbt]\centering
\label{fig_antikick}
{\scalebox{0.3}{\includegraphics{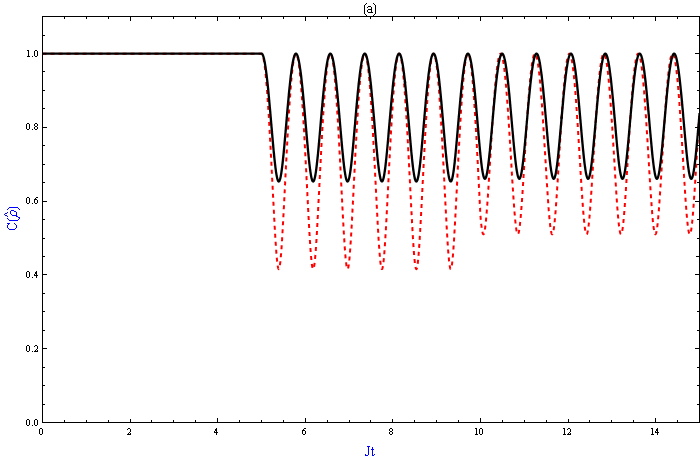}}}
{\scalebox{0.3}{\includegraphics{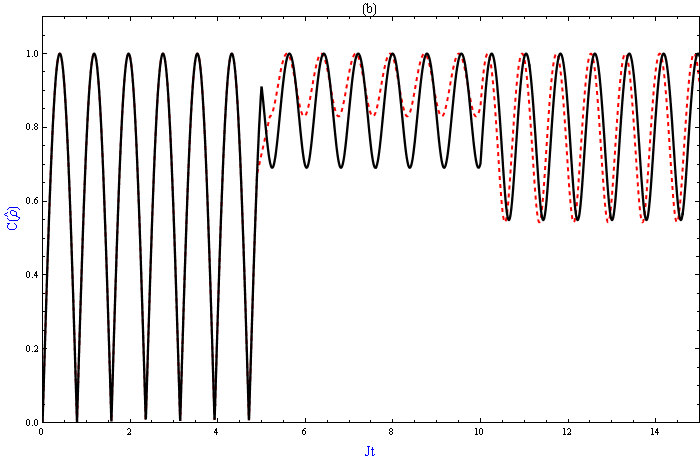}}}
\caption{Concurrence as a function of $Jt$ for an ideal positive kick applied at $T_1=5$ followed by an ideal negative kick at $T_2=10$ for the initial pure states $\left|\Phi(0)\right\rangle=\frac{1}{\sqrt{2}}(\left|10\right\rangle+\left|01\right\rangle)$~(a) and $\left|\Psi(0)\right\rangle=\left|01\right\rangle$~(b). The dashed lines correspond to $\alpha=2\beta$ and the solid lines to $\alpha=3\beta$.}
\end{figure}

The entanglement dynamics under the positive-negative kick sequence at times $t>T_2$ is obtained by using the expression Eq.~(\ref{kickakickpara}) in Eqs.~(\ref{con1}) and~(\ref{a1234}) and are displayed in Fig.~2(a) and~(b) for the initial Bell and separable states, respectively. The effect of the negative kick at $T_2$ is found to be opposite for the $\left|\Phi(0)\right\rangle=\frac{1}{\sqrt{2}}(\left|10\right\rangle+\left|01\right\rangle)$ and $\left|\Psi(0)\right\rangle=\left|01\right\rangle$ initial states; for the $\left|01\right\rangle$ state, the dependence of concurrence on the integrated magnetic strength vanishes while for the initial Bell state, amplitude of concurrence oscillations change with $\alpha$ and $\beta$. One peculiar result from Fig.~2(a) is the observation that the negative kick has no influence on the dynamics of concurrence for $\alpha=3\beta$ magnetic fields~(solid line in Fig.~2(a)). The positive-negative kick sequence also demonstrates the strong effect of time ordering on the entanglement dynamics. As mentioned before, based on the propagator for $\hat{T}\rightarrow 1$ case given by Eq.~(\ref{freeevolution}), the concurrence for Bell state is always equal to 1, while the concurrence for initially separable state has oscillations between 0 and 1 for the times $t>T_2$. On the other hand, as can be seen from Figs.~2(a) and~2(b), the time-ordered propagator leads to different results in the concurrence for both initial states.
\subsection{Two positive kicks}
To show the difference between positive and negative kicks applied after the first positive kick on the entanglement dynamics of two qubits, one may consider a sequence of two positive kicks applied at times $t=T_1$ and $t=T_2$, namely, $B_z^1(t)=\alpha\left(\delta(t-T_1)+\delta(t-T_2)\right), B_z^2(t)=\beta\left(\delta(t-T_1)+\delta(t-T_2)\right)$.  Following the procedure given in Eq.~(\ref{U^K}), one obtains the time evolution matrix Eq.~(\ref{kickakick}) for $t > T_2$ in the form as Eq.~(\ref{u0pulse}) with parameters
\begin{eqnarray}
\label{twokickpara}
y&=&e^{iJt},\nonumber\\
y_{1}&=&e^{2i(\alpha+\beta)},\nonumber\\
u&=&\cos\left(2Jt\right)\cos(\Delta)^2-\cos\left(2J(t-2 T_s)\right)\sin(\Delta)^2,\nonumber\\
v&=&\left(\cos(\zeta_1)+\cos(\zeta_2)\right)\sin(\Delta)\cos(\Delta),\nonumber\\
w&=&\left(\sin(\zeta_1)+\sin(\zeta_2)\right)\sin(\Delta)\cos(\Delta),\nonumber\\
z&=&-\sin\left(2Jt\right)\cos(\Delta)^2+\sin\left(2J(t-2 T_s)\right)\sin(\Delta)^2, 
\end{eqnarray}
where $\zeta_i =2J(t-2 T_i),\Delta=(\alpha-\beta)$ and $T_s=T_2-T_1$. Here the propagator  without time ordering can be calculated by replacing $\bar{B}_z^1 t \rightarrow 2 \alpha$ and $\bar{B}_z^2 t \rightarrow 2 \beta$ in Eq.~(\ref{u0pulse}) and note for $\alpha=\beta$ or $J=0$, $\hat{U}^K(t)-\hat{U}^0(t)$ vanishes, as expected according to Eqs.~(\ref{expanding}) and~(\ref{com}).
\begin{figure}[!hbt]\centering
\label{fig_3pkick}
{\scalebox{0.3}{\includegraphics{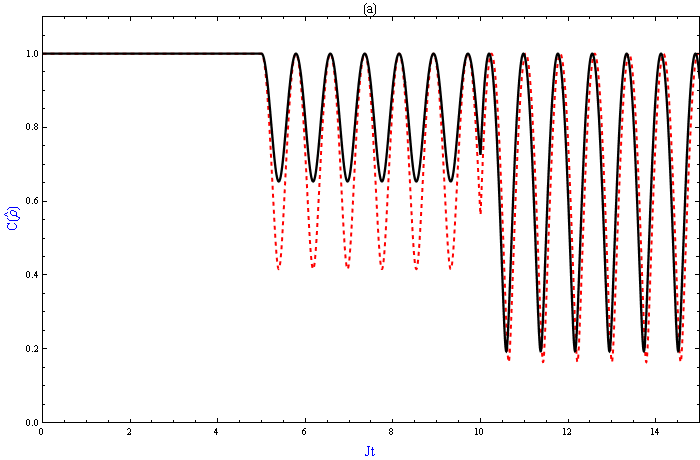}}}
{\scalebox{0.3}{\includegraphics{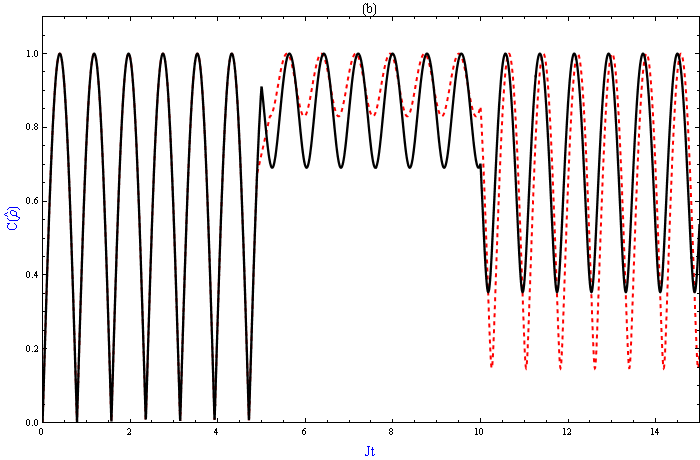}}}
\caption{Concurrence as a function of $Jt$ for a sequence of two ideal positive kicks applied at $T_1=5$ and $T_2=10$ for the initial pure states $\left|\Phi(0)\right\rangle=\frac{1}{\sqrt{2}}(\left|10\right\rangle+\left|01\right\rangle)$~(a) and $\left|\Psi(0)\right\rangle=\left|01\right\rangle$~(b). The dashed lines correspond to $\alpha=2\beta$ and the solid lines to $\alpha=3\beta$.}
\end{figure}

The effect of two consecutive positive kicks on the dynamics of concurrence for two qubits is displayed in Fig.~3(a) and~3(b) for the initial Bell state, $\frac{1}{\sqrt{2}}(\left|10\right\rangle+\left|01\right\rangle)$ and separable state, $\left|01\right\rangle$, respectively. Comparing the analytic expressions of the time-evolution operators for positive-negative and positive-positive kick sequences of Eq.~(\ref{kickakickpara}) and Eq.~(\ref{twokickpara}), respectively, along with the Fig.~2 and Fig.~3, the effect of the sign of the kicks in the sequence is to change the amplitude of the concurrence oscillations. The oscillation amplitude of $C(\hat{\rho})$ for the Bell state as well as separable state increases for the positive-positive sequence compared to that of positive-negative sequence of kicks. Also, $\alpha/\beta$ dependence of the amplitude is different as can be seen from a comparison of Fig.~2 and~3.

\subsection{Three and four positive kicks}
\label{multiplekick}
\begin{figure}[!hbt]\centering
\label{fig_3-4}
{\scalebox{0.3}{\includegraphics{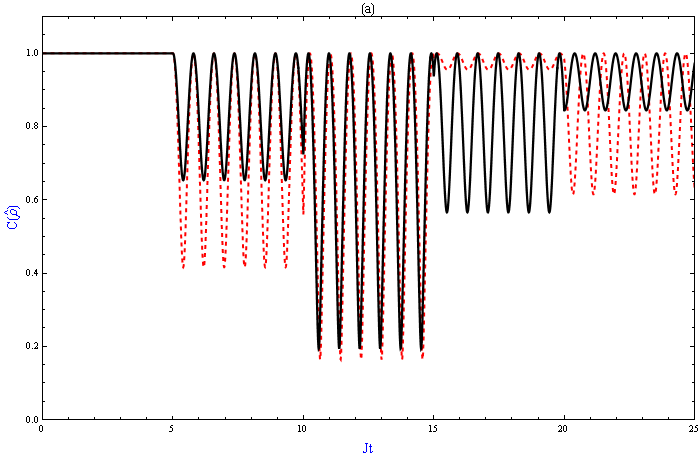}}}
{\scalebox{0.3}{\includegraphics{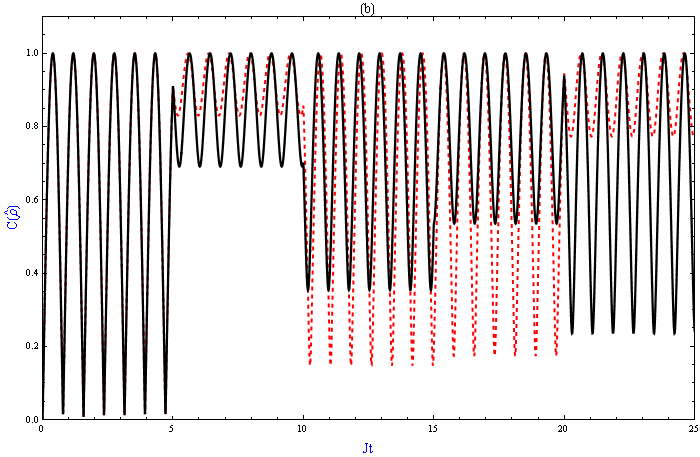}}}
\caption{Concurrence as a function of dimensionless time, $Jt$, for 4-successive ideal positive kicks for the initial pure states $\left|\Phi(0)\right\rangle=\frac{1}{\sqrt{2}}(\left|10\right\rangle+\left|01\right\rangle)$~(a) and $\left|\Psi(0)\right\rangle=\left|01\right\rangle$~(b). Here the dashed lines correspond to $\alpha=2\beta$ and the solid lines to $\alpha=3\beta$ and we take $T_1=5, T_2=10, T_3=15$ and $T_4=20$.}
\end{figure}
\begin{figure}[!hbt]\centering
\label{fig_kickcontour}
{\scalebox{0.3}{\includegraphics{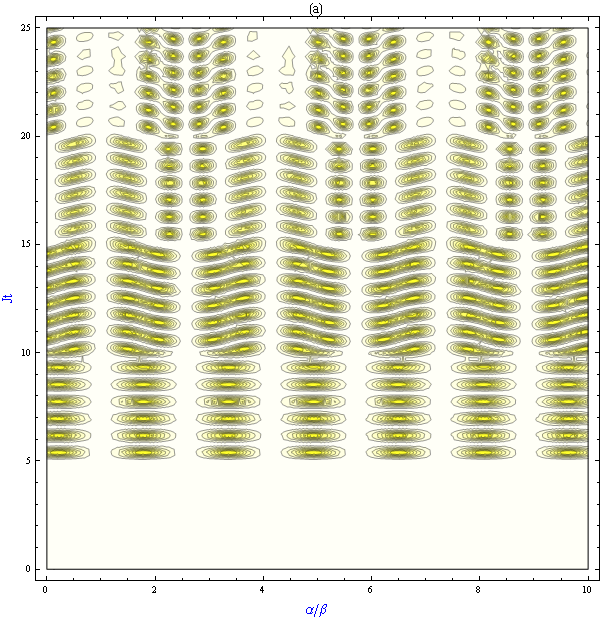}}}
{\scalebox{0.3}{\includegraphics{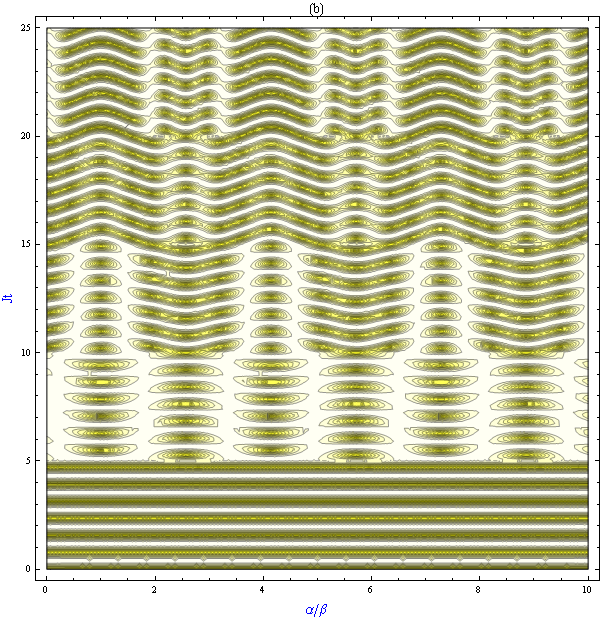}}}
\caption{(Colour online) The contour plot of concurrence versus $Jt$ and the ratio, $\alpha/\beta$, for the initial pure states $\left|\Phi(0)\right\rangle=\frac{1}{\sqrt{2}}(\left|10\right\rangle+\left|01\right\rangle)$~(a) and  $\left|\Psi(0)\right\rangle=\left|01\right\rangle$~(b). Here the contour plots include four ideal positive kicks applied at $T_1=5, T_2=10, T_3=15$ and $T_4=20$. (There are ten equidistant contours of concurrence in the plots between 0 (black) and 1 (white).)}
\end{figure}
One may consider a sequence of $n$-positive kicks applied  at times $t=T_1, t=T_2,...,t=T_n$, namely $B_z^1(t)=\displaystyle\sum_{i=1}^n\alpha\delta(t-T_i), B_z^2(t)=\displaystyle\sum_{i=1}^n \beta\delta(t-T_i)$. For example, following the procedure given in Eq.~(\ref{U^K}), one obtains the time evolution matrix for three positive kicks at times $t > T_3$ as:
\begin{eqnarray}
\hat U^{K}(t)&=&e^{-i\hat{H}_0(t-T_3)}e^{-i\int_{T_3 -\epsilon}^{T_3 +\epsilon}\hat{H}_{int}(t') dt'}e^{-i\hat{H}_0(T_3-T_2)}e^{-i\int_{T_2-\epsilon}^{T_2+\epsilon}\hat{H}_{int}(t') dt'}\nonumber\\
&\times& e^{-i\hat{H}_0(T_2-T_1)}e^{-i\int_{T_1-\epsilon}^{T_1+\epsilon}\hat{H}_{int}(t') dt'} e^{-i\hat{H}_0T_1},
\end{eqnarray}
in the same form as Eq.~(\ref{u0pulse}) with parameters 
\begin{eqnarray}
y&=& e^{iJt},\nonumber\\
y_{1}&=&e^{3i(\alpha+\beta)},\nonumber\\
u&=&\cos\left(2Jt\right)\cos(\Delta)^3-\sum_{\substack{i,j=1\\i<j}}^3 \cos \left(2J(t+2(T_i-T_j))\right)\cos(\Delta)\sin(\Delta)^2,\nonumber
\end{eqnarray}
\begin{eqnarray}\label{threekickspara}
v&=&\sum_{i=1}^3\cos(\zeta_i)\sin(\Delta)\cos (\Delta)^2-\cos\left(2J(t-2(T_1-T_2+T_3))\right)\sin(\Delta)^3,\nonumber\\
w&=&\sum_{i=1}^3\sin(\zeta_i)\sin (\Delta)\cos (\Delta)^2-\sin \left(2J(t-2(T_1-T_2+T_3))\right)\sin(\Delta)^3,\nonumber\\
z&=&-\sin \left(2Jt\right)\cos(\Delta)^3+\sum_{\substack{i,j=1\\i<j}}^3\sin\left(2J(t+2(T_i-T_j))\right)\cos(\Delta)\sin(\Delta)^2.\nonumber\\
\end{eqnarray}
Similarly, the time evolution matrix for four positive kicks at times $t>T_4$
\begin{eqnarray}
\hat{U}^{K}(t)&=&e^{-i\hat{H}_0(t-T_4)}e^{-i\int_{T_4-\epsilon}^{T_4+\epsilon}\hat{H}_{int}(t') dt'}e^{-i\hat{H}_0(T_4-T_3)}
e^{-i\int_{T_3-\epsilon}^{T_3+\epsilon}\hat{H}_{int}(t')dt'}e^{-i\hat{H}_0(T_3-T_2)}\nonumber\\
&\times&e^{-i\int_{T_2-\epsilon}^{T_2+\epsilon}\hat{H}_{int}(t')dt'}e^{-i\hat{H}_0(T_2-T_1)}e^{-i\int_{T_1-\epsilon}^{T_1+\epsilon}\hat{H}_{int}(t') dt'}e^{-i\hat{H}_0T_1},
\end{eqnarray}
with parameters specified in Eq.~(\ref{u0pulse})
\begin{eqnarray}
\label{fourkickspara}
y&=& e^{iJt},\nonumber\\
y_{1}&=&e^{4i(\alpha+\beta)},\nonumber\\
u&=&\cos\left(2Jt\right)\cos(\Delta)^4-\sum_{\substack{i,j=1\\i<j}}^4\cos\left(2J(t+2(T_i-T_j))\right)\cos(\Delta)^2\sin(\Delta)^2\nonumber\\
&+&\cos\left(2J(t+2 T_{1234})\right)\sin(\Delta)^4,\nonumber\\
v&=&\sum_{i=1}^4\cos(\zeta_i)\sin(\Delta)\cos(\Delta)^3-\sum_{\substack{i,j,k=1\\i<j<k}}^4\cos\left(2J(t-2(T_i-T_j+T_k))\right)\cos(\Delta)
\sin(\Delta)^3,\nonumber\\
w&=&\sum_{i=1}^4\sin(\zeta_i)\sin(\Delta)\cos(\Delta)^3-\sum_{\substack{i,j,k=1\\i<j<k}}^4\sin\left(2J(t-2(T_i-T_j+T_k))\right)\cos(\Delta) \sin(\Delta)^3,\nonumber\\
z&=&-\sin \left(2Jt\right)\cos(\Delta)^4+\sum_{\substack{i,j=1\\i<j}}^4\sin\left(2J(t+2(T_i-T_j))\right)\cos(\Delta)^2\sin(\Delta)^2\nonumber\\
&-&\sin\left(2J(t+2T_{1234})\right)\sin(\Delta)^4,
\end{eqnarray}
where $\zeta_i=2J(t-2 T_i), \Delta=(\alpha-\beta)$ and $T_{1234}=(T_1-T_2+T_3-T_4)$. Here the propagator  without time ordering can be calculated by replacing $\bar{B}_z^1t\rightarrow n\alpha$ and $\bar{B}_z^2t\rightarrow n\beta$~(here $n=3$ for three positive kicks and $n=4$ for four positive kicks) in Eq.~(\ref{u0pulse}) and note that for $\alpha=\beta$ or $J=0$, the time ordering effect defined as $\hat{U}^K(t)-\hat{U}^0(t)$ disappears after the field is active.

The dynamics of $C(\hat{\rho})$ under three and four positive kick sequence is shown in Figs.~4(a) and~4(b) for $\frac{1}{\sqrt{2}}(\left|10\right\rangle+\left|01\right\rangle)$ and  $\left|01\right\rangle$ initial states, respectively. The most important finding from these figures is that almost constant high entanglement can be obtained after the $3^{rd}$ kick for the $\frac{1}{\sqrt{2}}(\left|10\right\rangle+\left|01\right\rangle)$ state and after the $1^{st}$ and $4^{th}$ kicks for the $\left|01\right\rangle$ state at external magnetic field ratio of $\alpha/\beta=2$.

The integrated magnetic field strength dependence of the concurrence dynamics is shown in Fig.~5 where we display the contour plot of $C(\hat{\rho})$ as functions of $\alpha/\beta$ and $Jt$ for $\frac{1}{\sqrt{2}}(\left|10\right\rangle+\left|01\right\rangle)$ and $\left|01\right\rangle$ initial states. For the $\frac{1}{\sqrt{2}}(\left|10\right\rangle+\left|01\right\rangle)$ initial Bell state, the maximally entangled state is found to be unaffected by the external field for $\alpha/\beta=1$ and $\alpha/\beta\cong 2.5,4.25,5.75,7.25,8.75$. The  $\alpha/\beta$ periodicity of the maximum of $C(\hat{\rho})$ increases after each kick for the initial Bell state. The dependence of $C(\hat{\rho})$ on $\alpha/\beta$ for the initial separable state $\left|01\right\rangle$ is more complicated compared to the case of initial Bell state. The almost periodic structures exist also in Fig.~5(b); their periodicity changes after each kick, but it is not easy to obtain an expression for that change. Most importantly, the high entanglement regions, which are indicated in white in the contour plots have long life times for each positive kicks for the initial Bell state, while for the separable state they are distributed in a narrower area compared to the initial Bell state case and have long lifetimes only after $1^{st}$, $2^{nd}$ and $4^{th}$ kicks.
\section{Entanglement dynamics of qubits perturbed by a sequence of Gaussian pulses}
\label{gaussian}
Depending on the physical implementation of the qubit, it might be difficult to obtain an external field that can be considered as a kick. Instead a Gaussian pulse with finite width can be applied~(for example, half-cycle electromagnetic pulses with width near $\tau=1ps$ may be experimentally achievable~\cite{egp1,egp2}). Thus in this part, we will discuss  the entanglement dynamics of two qubits under the influence of  Gaussian pulses of the form $B_z^i(t)=\frac{\alpha_i}{\sqrt{\pi} \tau}e^{-\frac{(t-T_k)^2}{\tau^2}}$~($\alpha_{1,2}=\alpha,\beta$) centered at $T_k$ with width $\tau$. The dynamics of entanglement in the presence of time ordering for the initial pure states $\left|\Phi(0)\right\rangle=\frac{1}{\sqrt{2}}(\left|10\right\rangle+\left|01\right\rangle)$ and $\left|\Psi(0)\right\rangle=\left|01\right\rangle$ are evaluated by numerically integrating the corresponding equations in Eq.~(\ref{hamiltonian}) and using Eq.~(\ref{con1}). Here we will investigate how the entanglement depends on the pulse width $\tau$ by choosing a single pulse, a positive pulse followed by a negative pulse, and multiple positive pulses up to four  centered at times $T_1=5, T_2=10, T_3=15$ and $T_4=20$. For all pulse sequences we will consider two integrated magnetic strength regimes: $\alpha=2\beta$ and $\alpha=3\beta$ and for convenience we shall set $J=1$ and $\beta=1$. One should note that in the limit $\tau\rightarrow0$, the results of entanglement dynamics of kicked qubits in the presence of time ordering should be the same that are analyzed in the previous section.

\subsection{Single Pulse}
\label{singlepulse}
\begin{figure}[ht]\centering
\label{fig_singlegaussian}
{\scalebox{0.25}{\includegraphics{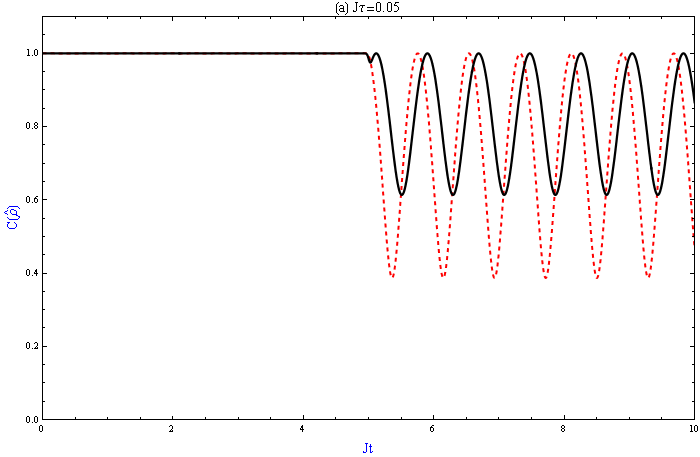}}}
{\scalebox{0.25}{\includegraphics{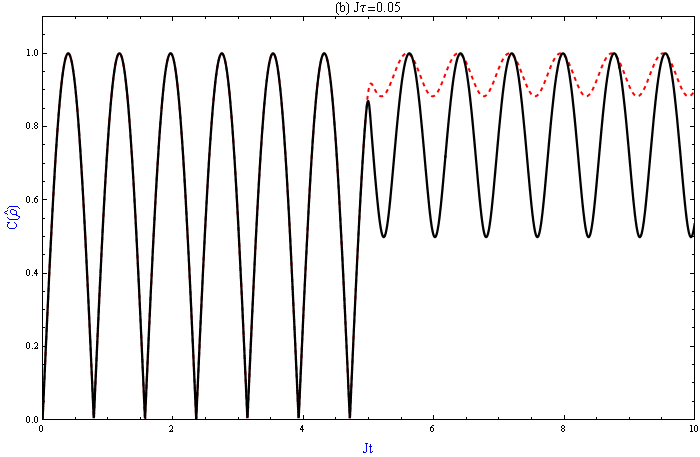}}}

{\scalebox{0.25}{\includegraphics{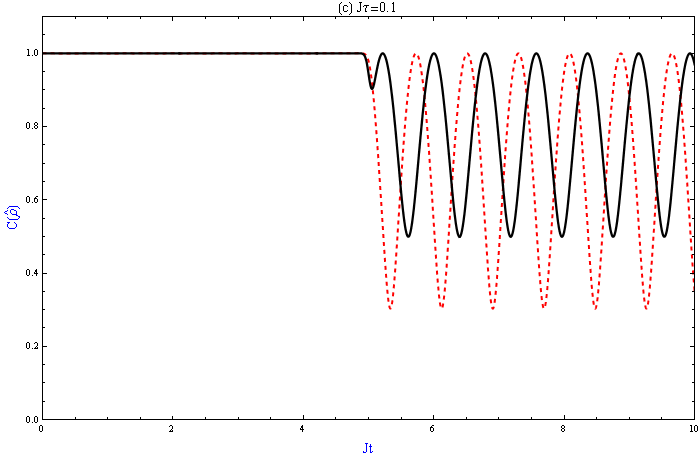}}}
{\scalebox{0.25}{\includegraphics{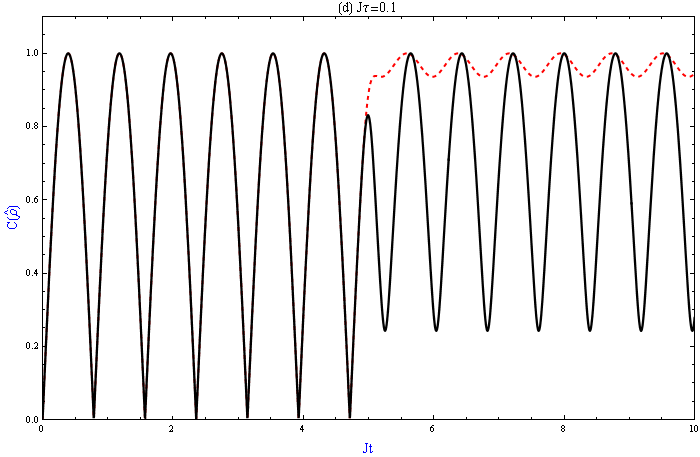}}}

{\scalebox{0.25}{\includegraphics{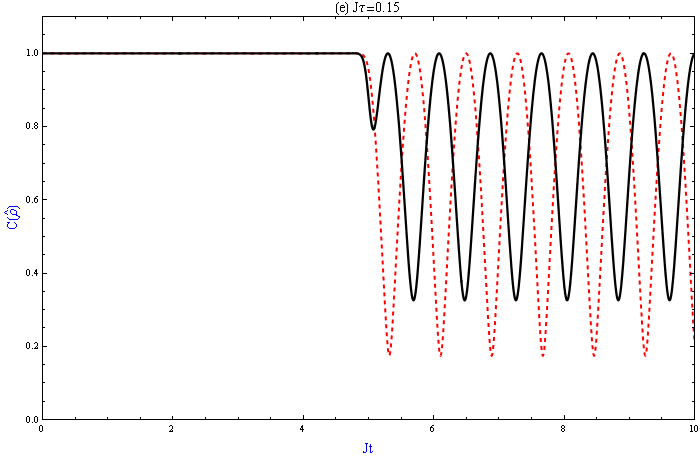}}}
{\scalebox{0.25}{\includegraphics{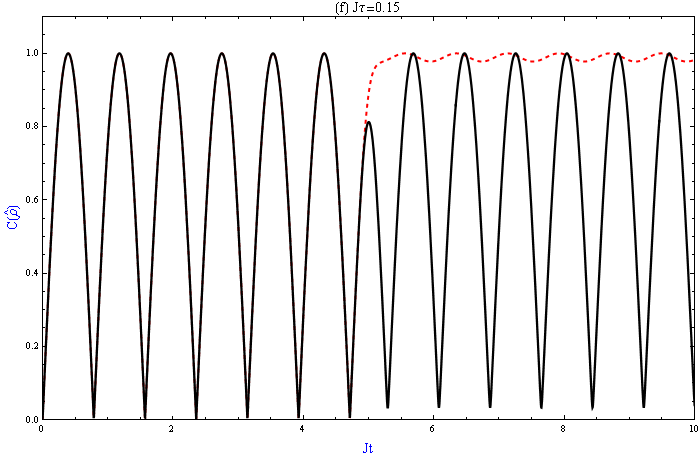}}}

{\scalebox{0.25}{\includegraphics{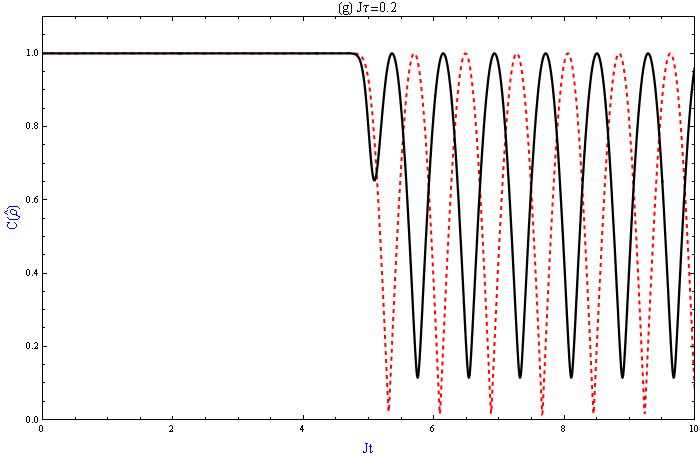}}}
{\scalebox{0.25}{\includegraphics{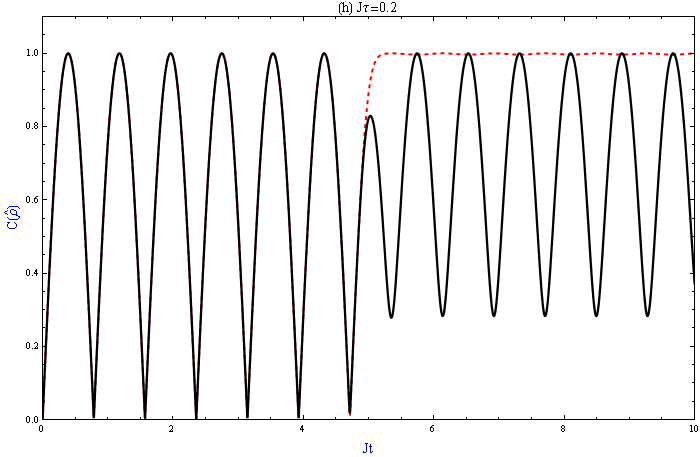}}}
\caption{Concurrence as a function of $Jt$ for a single Gaussian pulse with width $\tau$  for the initial pure states $ \left|\Phi(0)\right\rangle=\frac{1}{\sqrt{2}}(\left|10\right\rangle+\left|01\right\rangle)$~(a),~(c),~(e) and~(g)  and $\left|\Psi(0)\right\rangle=\left|01\right\rangle$~(b),~(d),~(f) and~(h). The dashed lines correspond to $\alpha=2\beta$ and the solid lines to $\alpha=3\beta$. Here we assume four dimensionless pulse width as: $J\tau=0.05,0.1,0.15,0.2$.}
\end{figure}

In Fig.~6, we show the results of a calculation of the concurrence for the initial pure states $\left|\Phi(0)\right\rangle=\frac{1}{\sqrt{2}}(\left|10\right\rangle+\left|01\right\rangle)$ and  $\left|\Psi(0)\right\rangle=\left|01\right\rangle$ when strongly perturbed by a single Gaussian pulse centered at $t=T_1$ with width $\tau$. According to Schr\"{o}dinger equation~(\ref{hamiltonian}) of the system considered here, the expansion coefficients $a_1(t)$ and $a_4(t)$ evolve independently and $a_2(t)$ and $a_3(t)$ obey a first-order coupled differential equation set, for example for a single pulse, as:
\begin{eqnarray} 
\label{num1} 
i\dot{a}_2(t)&=&\left(-J-\frac{(\alpha-\beta)}{\sqrt{\pi}\tau}e^{-\frac{(t-T_1)^2}{\tau^2}}\right)a_2(t)+2 J a_3(t),\nonumber\\
i\dot{a}_3(t)&=&\left(-J+\frac{(\alpha-\beta)}{\sqrt{\pi}\tau}e^{-\frac{(t-T_1)^2}{\tau^2}}\right)a_3(t)+2 J a_2(t),
\end{eqnarray}
which are solved numerically by using a fourth order Runge-Kutta algorithm. The most important observation from Fig.~6 is the existence of almost constant high concurrence for the initially separable state at $\alpha=2\beta$ integrated magnetic strength and high width Gaussian pulse, while the entanglement continues to have high amplitude oscillations for $\alpha=3\beta$; its value for $\alpha=2\beta$ is almost constant at around 1 for $J\tau$; the dimensionless pulse width greater than 0.15. On the contrary for the initial Bell state, the oscillation amplitude of $C(\hat{\rho})$ increases with the $J\tau$ of the pulse for each magnetic ratio~($\alpha/\beta=2$ and $\alpha/\beta=3$).
\subsection{Positive-negative and positive-positive pulse sequence}
\label{antipulse}
\begin{figure}[ht]\centering
\label{fig_psgaussian}
{\scalebox{0.25}{\includegraphics{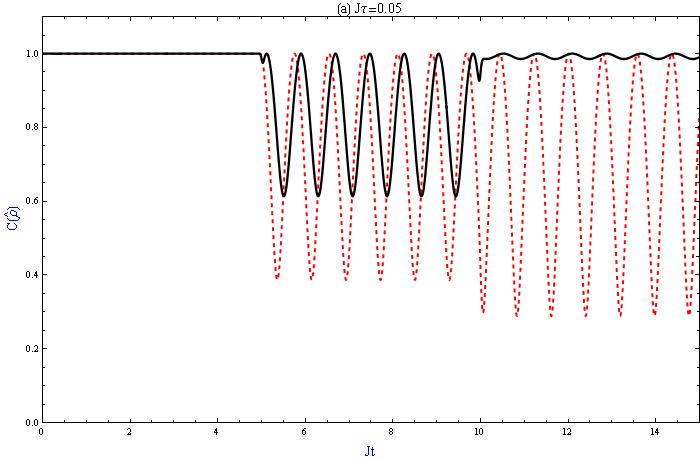}}}
{\scalebox{0.25}{\includegraphics{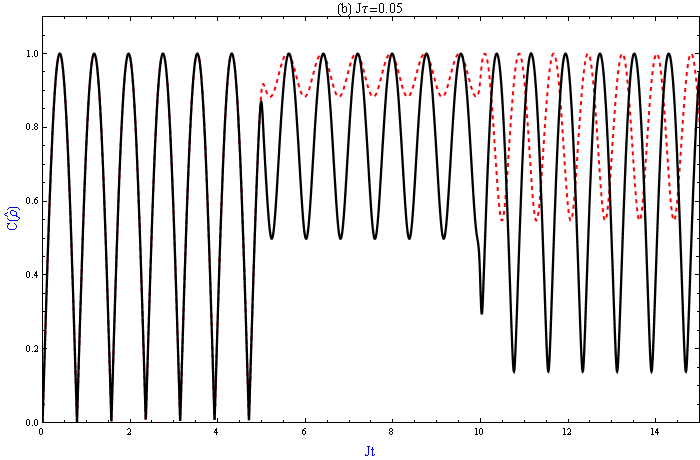}}}

{\scalebox{0.25}{\includegraphics{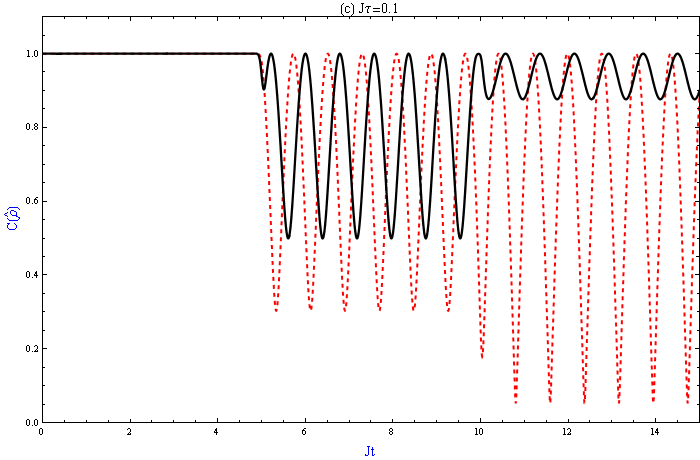}}}
{\scalebox{0.25}{\includegraphics{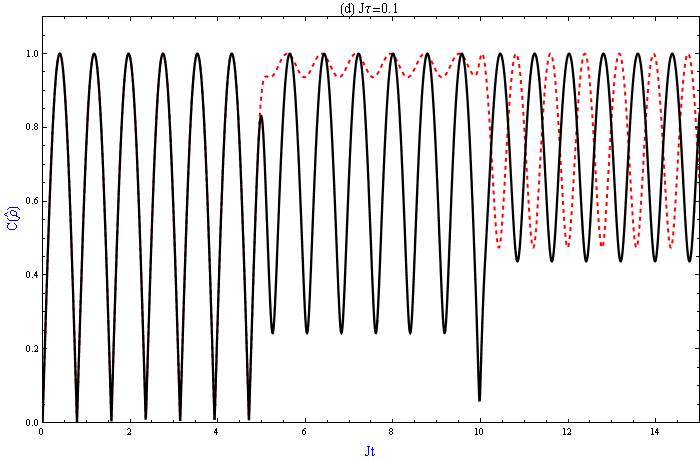}}}

{\scalebox{0.25}{\includegraphics{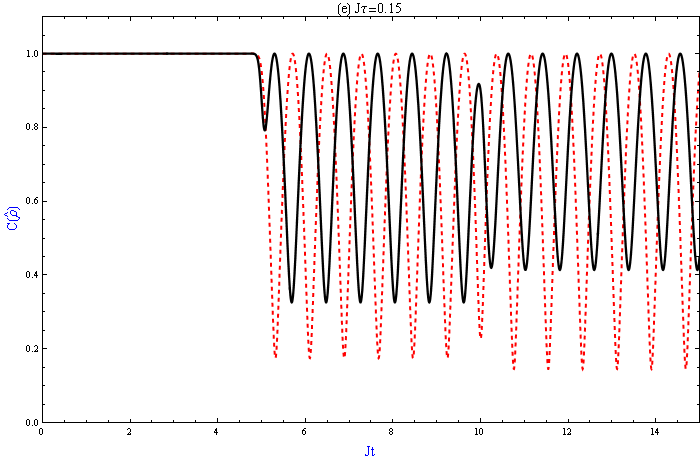}}}
{\scalebox{0.25}{\includegraphics{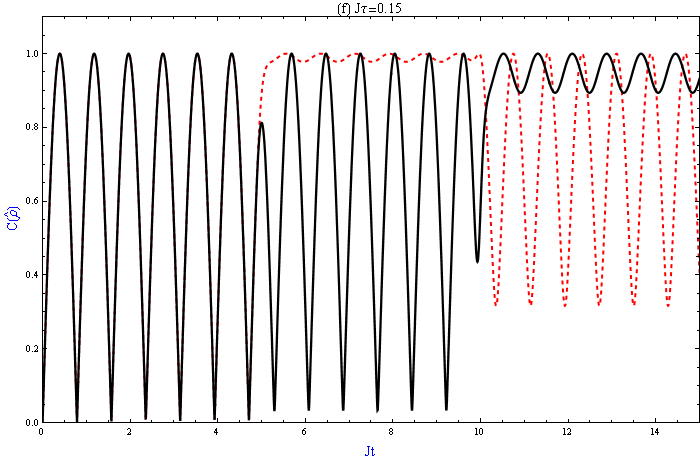}}}

{\scalebox{0.25}{\includegraphics{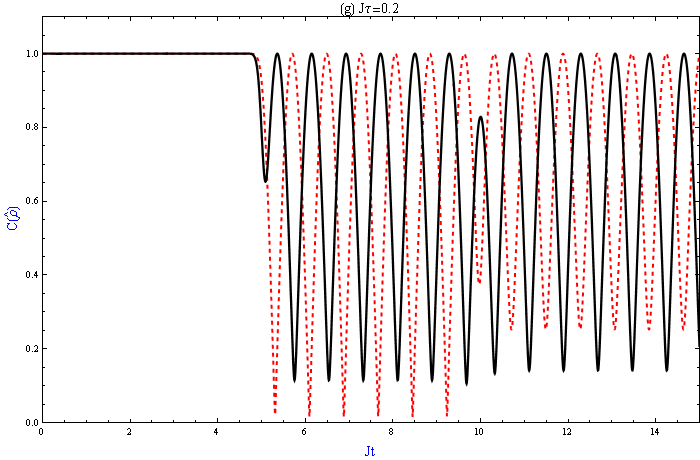}}}
{\scalebox{0.25}{\includegraphics{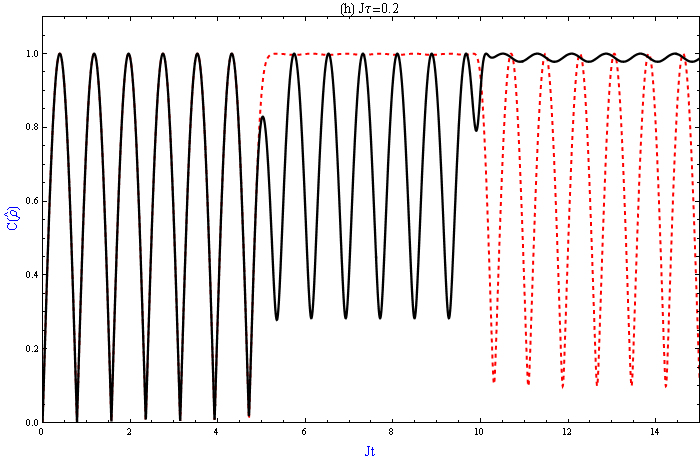}}}
\caption{Concurrence as a function of $Jt$ for a positive followed by a negative Gaussian pulses having the same width $\tau$ for the initial pure states $\left|\Phi(0)\right\rangle=\frac{1}{\sqrt{2}}(\left|10\right\rangle+\left|01\right\rangle)$~(a),~(c),~(e) and~(g) and $\left|\Psi(0)\right\rangle=\left|01\right\rangle$~(b),~(d),~(f) and~(h). The dashed lines correspond to $\alpha=2\beta$ and the solid lines to $\alpha=3\beta$. Here we assume four dimensionless pulse width as: $J\tau=0.05,0.1,0.15,0.2$.}
\end{figure}

In Fig.~7 and~8, we show the results of a calculation for the concurrence for the initial pure states $\left|\Phi(0)\right\rangle=\frac{1}{\sqrt{2}}(\left|10 \right\rangle+\left|01\right\rangle)$ and  $\left|\Psi(0)\right\rangle=\left|01\right\rangle$ when strongly perturbed by a single Gaussian pulse centered at $t=T_1$ followed by a negative or positive Gaussian pulse centered at $t=T_2$ with the same width $\tau$. For the double pulse sequence $a_2(t)$ and $a_3(t)$ obey the coupled equations:
\begin{eqnarray} 
\label{num2} 
i\dot{a}_2(t)&=&\left(-J-\frac{(\alpha-\beta)}{\sqrt{\pi}\tau}(e^{-\frac{(t-T_1)^2}{\tau^2}}\pm e^{-\frac{(t-T_2)^2}{\tau^2}})\right)a_2(t)+2Ja_3(t),\nonumber\\
i\dot{a}_3(t)&=&\left(-J+\frac{(\alpha-\beta)}{\sqrt{\pi}\tau}(e^{-\frac{(t-T_1)^2}{\tau^2}}\pm e^{-\frac{(t-T_2)^2}{\tau^2}})\right)a_3(t)+2Ja_2(t),
\end{eqnarray}
where $+$ sign in the $\pm$ on the right-hand side is for positive-positive pulse sequence, while $-$ sign is for positive-negative pulse sequence.

\label{twopulses}
\begin{figure}[ht]\centering
\label{fig_twogaussian}
{\scalebox{0.25}{\includegraphics{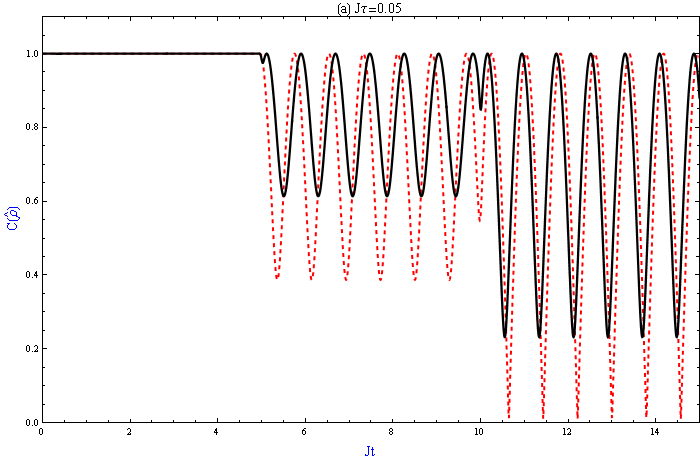}}}
{\scalebox{0.25}{\includegraphics{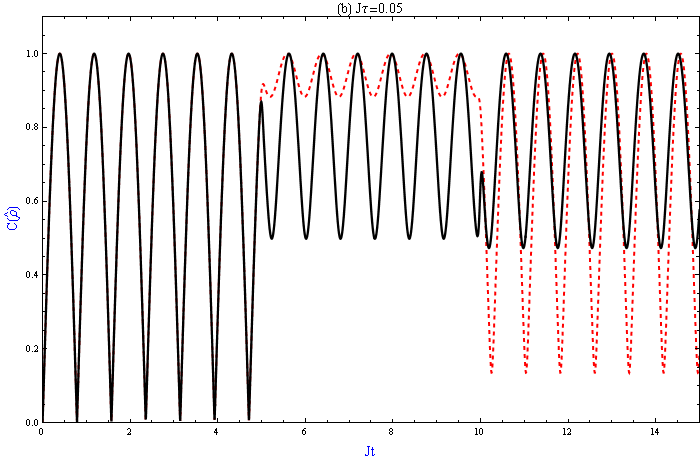}}}

{\scalebox{0.25}{\includegraphics{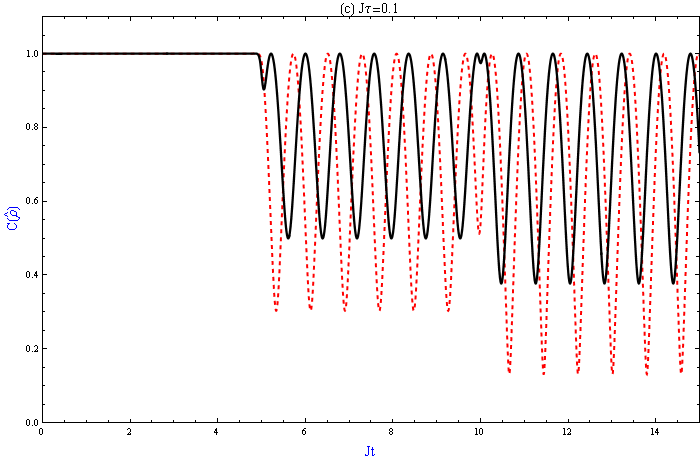}}}
{\scalebox{0.25}{\includegraphics{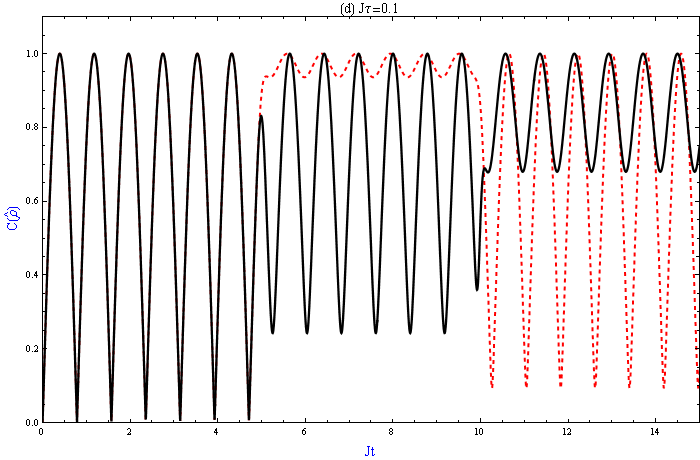}}}

{\scalebox{0.25}{\includegraphics{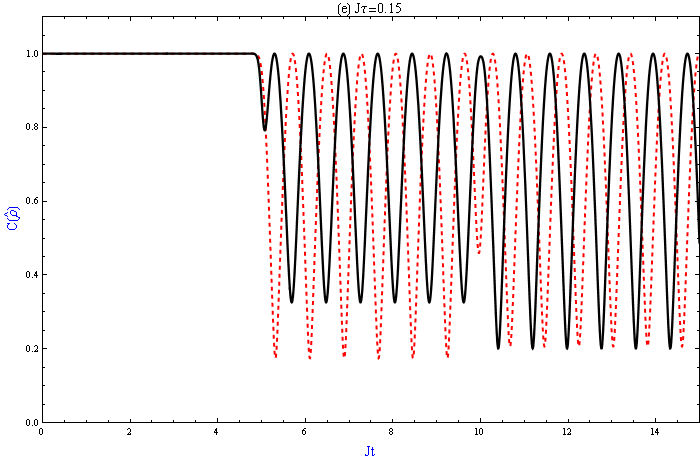}}}
{\scalebox{0.25}{\includegraphics{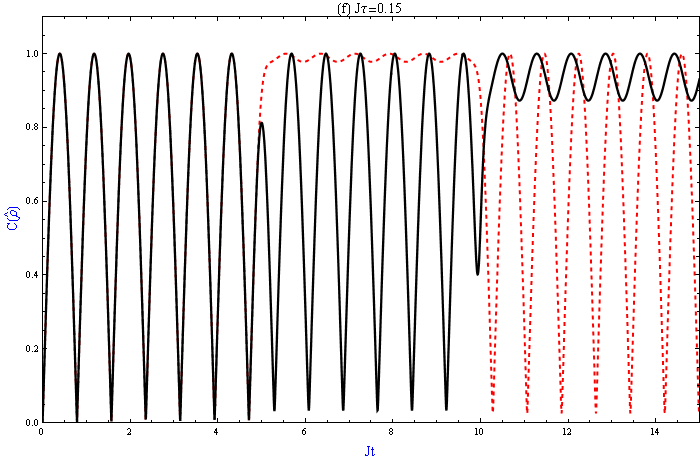}}}

{\scalebox{0.25}{\includegraphics{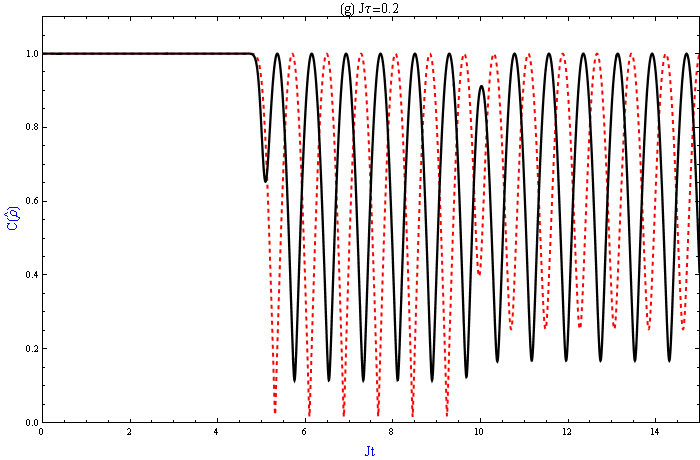}}}
{\scalebox{0.25}{\includegraphics{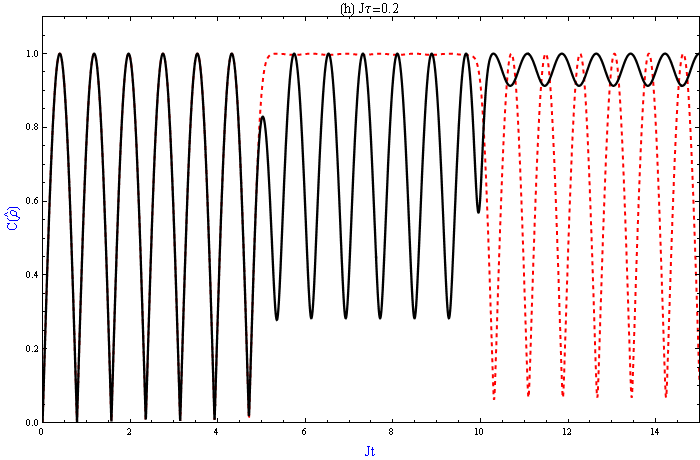}}}
\caption{Concurrence as a function of $Jt$ for a sequence of two positive Gaussian pulses having the same witdh $\tau$ for the initial pure states $ \left|\Phi(0)\right\rangle=\frac{1}{\sqrt{2}}(\left|10\right\rangle+\left|01\right\rangle)$~(a),~(c),~(e) and~(g) and $\left|\Psi(0)\right\rangle=\left|01\right\rangle$~(b),~(d),~(f) and~(h). The dashed lines correspond to $\alpha=2\beta$ and the solid lines to $\alpha=3\beta$. Here we assume four dimensionless pulse width as: $J\tau=0.05,0.1,0.15,0.2$.}
\end{figure}

The concurrence dynamics for positive-negative and positive-positive pulse sequences obtained from the numerical solutions of Eq.~(\ref{num2}) are displayed in Figs.~7 and~8, respectively for the initial Bell and the separable states at different dimensionless pulse width values~($J\tau=0.05,0.10,0.15,0.20$). For the initial Bell state under positive-negative sequence, the most important effect of the pulse width seems to be an increase in the $C(\hat{\rho})$ oscillation amplitudes for $\alpha=3\beta$ at times $t>T_2$. The almost constant high entanglement can be obtained for the $\left|01\right\rangle$ initial state for positive-negative pulse sequence as can be seen from Fig.~7(h) for $\alpha=3\beta$ and $t>T_2$. One peculiarity of this figure is that high entanglement is obtained for $\alpha=2\beta$ after the first pulse, while it is obtained for $\alpha=3\beta$ after the second pulse. For the positive-positive Gaussian pulse sequence, the difference from positive-negative sequence becomes small as the width of the pulse gets larger as can be deduced from a comparison of Figs.~7 and~8. On the other hand, for a small pulse width, the difference is significant. For example for $J\tau=0.05$ and $\alpha=3\beta$, the initial Bell state has nearly constant entanglement around 1~(see Fig.~7(a)) for positive-negative pulse sequence after negative pulse, while it oscillates between 1 and 0.25 for positive-positive pulse sequence~(see Fig.~8(a)).
\subsection{A sequence of four positive pulses}
The effect of integrated magnetic strength and the pulse width on the dynamics of concurrence for two qubits perturbed by a sequence of four positive Gaussian pulses is displayed in Fig.~9(a)-(h) for initial Bell state, $\left|\Phi(0)\right\rangle=\frac{1}{\sqrt{2}}(\left|10\right\rangle+\left|01\right\rangle)$, and separable state, $\left|\Psi(0)\right\rangle=\left|01\right\rangle$. For this four positive pulse sequence the concurrence may be calculated by using the numerical solutions of the coupled equations:
\begin{eqnarray} 
\label{num4} 
i\dot{a}_2(t)&=&\left(-J-\frac{(\alpha-\beta)}{\sqrt{\pi}\tau}\sum_{i=1}^4e^{-\frac{(t-T_i)^2}{\tau^2}}\right)a_2(t)+2 J a_3(t),\nonumber\\
i\dot{a}_3(t)&=&\left(-J+\frac{(\alpha-\beta)}{\sqrt{\pi}\tau}\sum_{i=1}^4e^{-\frac{(t-T_i)^2}{\tau^2}}\right)a_3(t)+2 J a_2(t).
\end{eqnarray}
These figures can be compared with Figs.~5(a) and~5(b) to discern the effect of the pulse width. As the pulse gets wider, the $\alpha/\beta$ dependent oscillatory structures in the figure coalesce to produce non-periodic structures, especially after third and fourth pulses. The high entanglement regions, which are shown in white, still can have long lifetimes, as indicated by white straight perpendicular sections in the contour plots of Fig.~9. In the case of ideal kick, the maximally entangled state is found to be unaffected by the external field for $\alpha/\beta=1$ and $\alpha/\beta\cong 2.5,4.25,5.75,7.25,8.75$~(see Fig.~5(a)). Comparing Fig.~5(a) with Figs.~9(a),~(c),~(e) and~(g), the initial Bell state is found to be unperturbed by the highly wider Gaussian pulses if and only if $\alpha/\beta=1$; especially seen obviously for the dimensionless pulse width greater than 0.15. One should note that this is one of the conditions in which time ordering effects vanishes.
\begin{figure}[ht]\centering
\label{fig_fourgaussiancontour}
{\scalebox{0.20}{\includegraphics{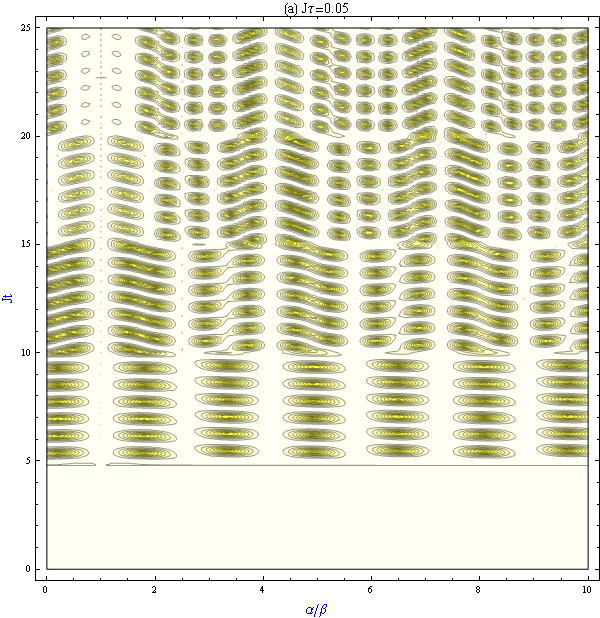}}}
{\scalebox{0.20}{\includegraphics{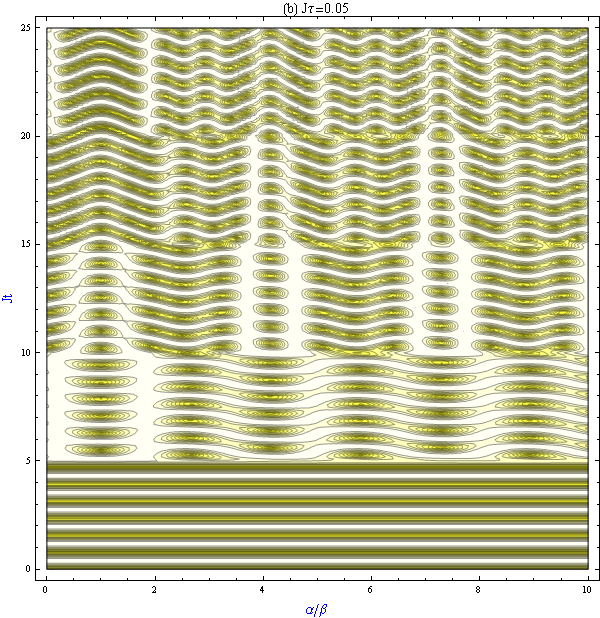}}}

{\scalebox{0.20}{\includegraphics{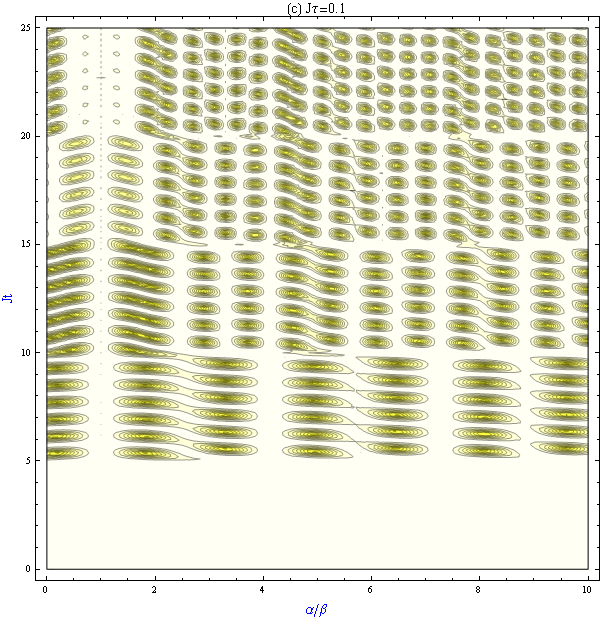}}}
{\scalebox{0.20}{\includegraphics{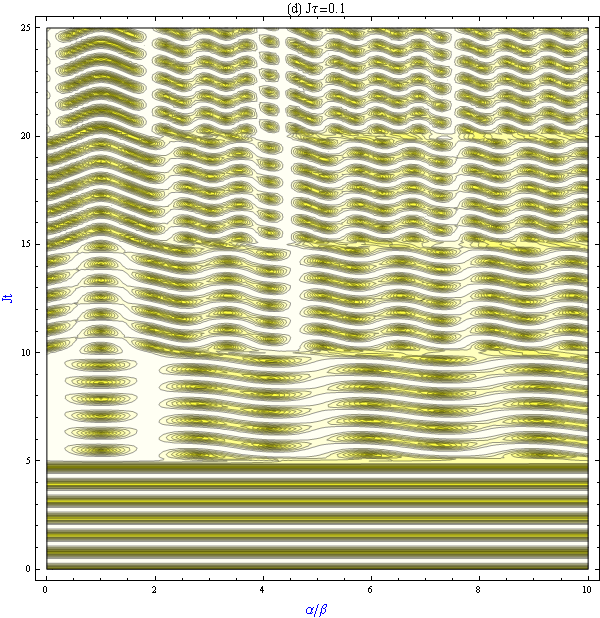}}}

{\scalebox{0.20}{\includegraphics{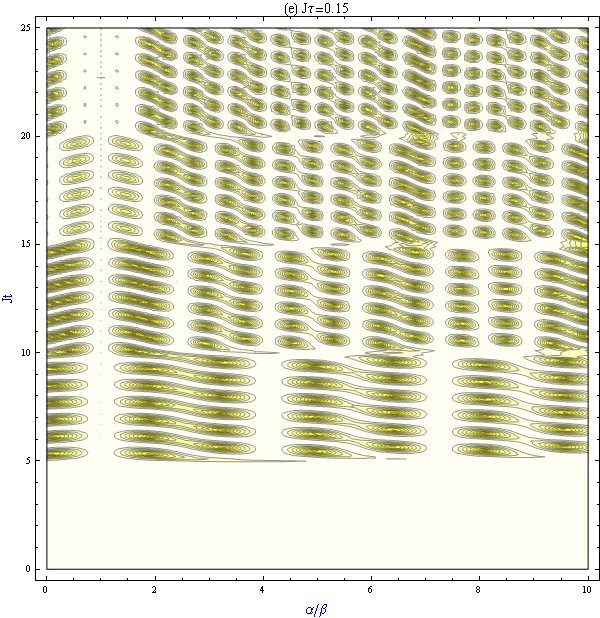}}}
{\scalebox{0.20}{\includegraphics{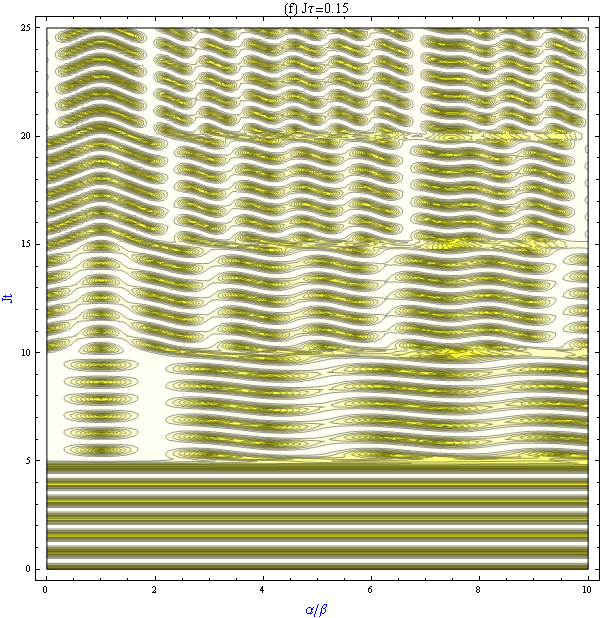}}}

{\scalebox{0.20}{\includegraphics{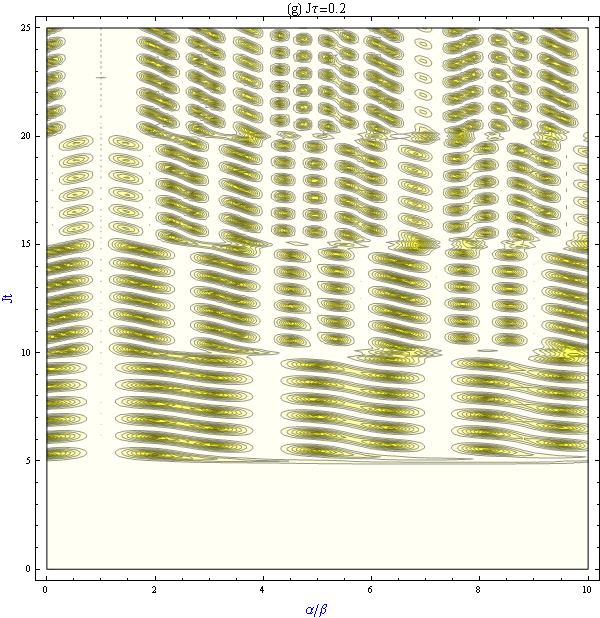}}}
{\scalebox{0.20}{\includegraphics{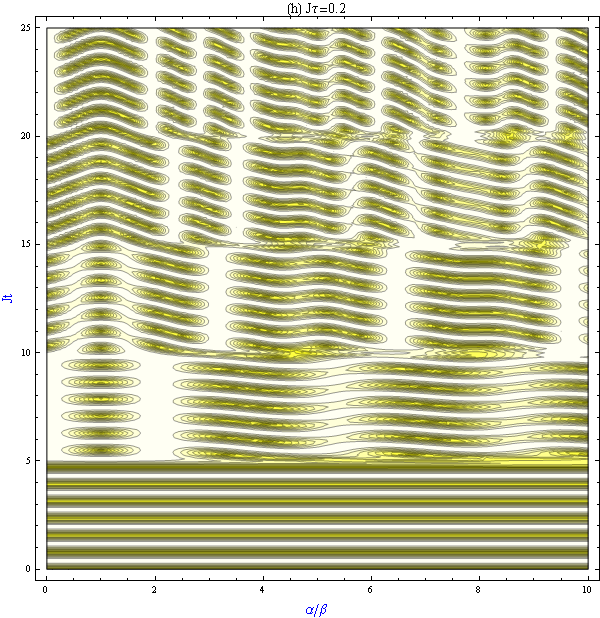}}}
\caption{(Colour online) The contour plot of concurrence versus $Jt$ and $\alpha/\beta$, for a sequence of four positive Gaussian pulses of width $\tau$ for the initial pure states $\left|\Phi(0)\right\rangle=\frac{1}{\sqrt{2}}(\left|10\right\rangle+\left|01\right\rangle)$~(a),~(c),~(e) and~(g) and $\left|\Psi(0)\right\rangle=\left|01\right\rangle$~(b),~(d),~(f) and~(h). Here we assume four dimensionless pulse width as: $J\tau=0.05,0.1,0.15,0.2$. (There are ten equidistant contours of concurrence in the plots between 0 (black) and 1 (white).) }
\end{figure}

The last point we want to emphasize that what happens the entanglement dynamics between two qubits if the time ordering effect vanishes. As mentioned before the time ordering effect vanishes for the special cases either $\alpha=\beta$ or $J=0$. From corresponding equations it can be noted that for the case $\alpha=\beta$, the concurrence function for the initial state  $\left|\Phi(0)\right\rangle=\frac{1}{\sqrt{2}}(\left|10\right\rangle+\left|10\right\rangle)$ is equal to $1$, while  for the initial state $\left|\Psi(0)\right\rangle=\left|01\right\rangle$ is equal to $\left|\sin\left(4Jt\right)\right|$ and unaffected by the sequence of the kicks or Gaussian pulses. For the other case, $J=0$, under the influence of kick or Gaussian pulse sequences the concurrence for the initial Bell state is always equal to 1, while the separable state remains separable at any time. These show an important fact that since the time ordering provides a connection between interactions at different times, it is responsible from the nonlocal correlations between the qubits in time.

\section{Conclusion}
\label{conc}
We have investigated the dynamics of entanglement for two qubits that interact with each other via Heisenberg XXX-type interaction under a time-dependent external magnetic field. Initial state of the system is considered to be pure Bell or separable states. The main aim of the study was to investigate the controllability of the entanglement with a sequence of pulse or kick type external fields.

The effect of time ordering in the dynamics of concurrence is found to be important; concurrence calculated when the time ordering is neglected is found to be completely different than when it is taken into account. Time-dependent concurrence obtained after one, two, three and four kicks at different magnetic field strengths indicate that one can employ carefully chosen kick or kick sequences to produce high entanglement between two initially non-entangled qubits.

We have also considered the effect of the pulse width of the external field on the entanglement dynamics by modelling the external field as a Gaussian pulse or a sequence of Gaussian pulses. Increasing the width of the pulse is found to enhance the control of high and steady entanglement. 

One should note that the external control field considered in the present study acts on both of the qubits at the same time. It might be possible to use pulse sequences acting an individual qubits at different times to obtain a better control of entanglement.


\section*{References}
\begin{thebibliography}{99}

\bibitem{nc} Nielson M A and Chuang I L 2000 Quantum Computation and Quantum Information, (Cambridge: Cambridge University Press)

\bibitem{bennett} Bennett C H, DiVincenzo D P, Smolin J A and Wootters W K 1996 {\it Phys. Rev. A} {\bf 54} 3824

\bibitem{ekert} Ekert A K 1991 {\it Phys. Rev. Lett.} {\bf 67} 661

\bibitem{hbbs} Heule R , Bruder C, Burgarth D and Stojanovic V M arXiv:quant-ph/1007.2572

\bibitem{cmcfmgs} Caneva T, Murphy M, Calarco T, Fazio F, Montangero S, Giovannetti V and Santoro G E 2009 {\it Phys. Rev. Lett.} {\bf 103} 240501

\bibitem{wl} Wu L A, Lidar L A and Friesen M 2004  {\it Phys. Rev. Lett.} {\bf 93} 030501

\bibitem{levy} Levy J 2002 {\it Phys. Rev. Lett.} {\bf 89}  147902

\bibitem{ms} Malinovsky V S and Sola I R 2004 {\it Phys. Rev. Lett.} {\bf 93} 190502

\bibitem{zmf} Sadiek G, Lashin E I and Abdalla M S 2009 {\it Physica B} {\bf 404} 1719

\bibitem{wbsb} Wang X, Bayat A, Schirmer S G and Bose S 2010 {\it Phys. Rev. A} {\bf 81}  032312

\bibitem{abliz} Abliz A, Gao H J, Xie X C, Wu Y S and Liu W M 2006 {\it Phys. Rev. A} {\bf 74}  052105

\bibitem{wang} Wang X 2001 {\it Phys. Rev. A} {\bf 64} 012313

\bibitem{sb} Sainz I, Burlak G and Klimov A B, arXiv:quant-ph/1008.2784

\bibitem{iabdlss} Imamoglu A, Awschalom D D, Burkard G, DiVicenzo D P, Loss D, Sherwin M and Small A 1999 {\it Phys. Rev. Lett.} {\bf 83} 4204

\bibitem{kaplan} Kaplan L, Shakov K K, Chalastaras A, Maggio M, Burin A L and McGuire J H 2004 {\it Phys. Rev. A} {\bf 70} 063401

\bibitem{shakov} Shakov K K, McGuire J H, Kaplan L, Uskov D and Chalastaras A 2006 {\it J. Phys. B: At. Mol. Opt. Phys.} {\bf 39}  1361

\bibitem{jhm} Jones J A, Hansen R A and Mosca M 1998 {\it  J. Magn. Reson.} {\bf 138} 353

\bibitem{vsbysc} Vandersypen L M K, Steffen M, Breyta G, Yannoni C S, Sherwood M H and Chuang I L 2001 {\it Nature} {\bf 414} 883

\bibitem{slichter} Slichter C P 1996, Principles of Magnetic Resonance, (Berlin: Springer)

\bibitem{krgt} Kosloff R, Hammerich A D and Tannor D 1992 {\it Phys. Rev. Lett.} {\bf 69}  2172

\bibitem{swr} Shi S, Woody A and Rabitz H 1988 {\it J. Chem. Phys.} {\bf 88} 6870

\bibitem{pk} Palao J and Kosloff R 2002 {\it Phys. Rev. Lett.} {\bf 89} 188301

\bibitem{rau} Rau A R P 1998 {\it Phys. Rev. Lett.} {\bf 81} 4785

\bibitem{rsu} Rau A R P, Selvaraj G and Uskov D 2005 {\it Phys. Rev. A} {\bf 71} 062316

\bibitem{dysont} Godunov A L and McGuire J H 2001 {\it J.~Phys.~B} {\bf 34} 223

\bibitem{zlt} Zhao H Z, Lu Z H and Thomas J E 1997 {\it Phys. Rev. Lett.} {\bf 79}  613

\bibitem{mbhgm} Merabet H, Bruch R, Hanni J, Godunov A L and McGuire J H 2002 {\it Phys. Rev. A} {\bf 65} 010703

\bibitem{magnus} Magnus W 1954 {\it Commun. Pure Appl. Math.} {\bf 7}  649

\bibitem{wcon} Wootters W K 1998 {\it Phys. Rev. Lett.} {\bf 80} 2245

\bibitem{egp1} Jones R R, You D and Bucksbaum P H 1993 {\it Phys. Rev. Lett.} {\bf 70} 1236

\bibitem{egp2} Abique A M and Berakdar B 2004 {\it Appl. Phys. Lett.} {\bf 84} 2346

\end{thebibliography}
\end{document}